\documentclass[a4paper,fleqn,usenatbib,useAMS]{mnras}

\usepackage{graphicx}
\usepackage{amsmath}
\usepackage{amssymb}
\usepackage{bm}
\usepackage{pdflscape}
\usepackage{float}
\usepackage[T1]{fontenc}
\usepackage{ae,aecompl}
\usepackage{color}
\usepackage{caption}

\pdfoutput=1 

\DeclareMathOperator{\sgn}{sgn}

\title[An Eccentric Path to Tidal Disruption]{Tidal Dissipation and Evolution of White Dwarfs Around Massive Black Holes: An Eccentric Path to Tidal Disruption}

\author[M. Vick, D. Lai and J. Fuller]{Michelle Vick$^{1}$, Dong Lai$^{1}$, Jim Fuller$^{2,3}$\\$^{1}$Cornell Center for Astrophysics and Planetary Science, Department of Astronomy, Cornell University, Ithaca, NY 14853, USA\\$^{2}$TAPIR, Walter Burke Institute for Theoretical Physics, Mailcode 350-17, Caltech, Pasadena, CA 91125, USA\\$^{3}$Kavli Institute for Theoretical Physics, Kohn Hall, University of California, Santa Barbara, CA 93106, USA}

\pubyear{2016}

\begin{document}

\label{firstpage}
\pagerange{\pageref{firstpage}--\pageref{lastpage}}
\maketitle

%%%%%%%%%%%%%%%%%%%%%%%%%%%%%%%
\begin{abstract}
%%%%%%%%%%%%%%%%%%%%%%%%%%%%%%%
A white dwarf (WD) captured into a high-eccentricity orbit around a
massive black hole (MBH) may undergo many pericenter passages before
tidal disruption. During these passages, the tidal potential of the
MBH excites internal oscillations or waves in the WD, and the dissipation of
these oscillations can significantly influence the physical properties of the
WD prior to its disruption. We calculate the amplitude of the tidally
excited gravity (buoyancy) waves in the WD as a function of the pericenter distance and
eccentricity for realistic WD models, under the assumption that these
outgoing gravity waves are efficiently dissipated in outer layers
of the WD by non-linear effects or radiative damping. We obtain fitting
formulae for the tidal energy and angular momentum transfer rates as
well as the tidal heating rate. We find that these dynamical tides are
much weaker than gravitational radiation in driving the orbital decay
of the WD-MBH binary, and they are also inefficient in changing the WD
spin during the orbital evolution.  Incorporating our computed tidal
dissipation rate into a MESA-based WD evolution code, we find that tidal
heating can lead to appreciable brightening of the WD and 
may induce runaway fusion in the hydrogen envelope well 
before the WD undergoes tidal disruption.

\end{abstract}

%%%%%%%%%%%%%%%%%%%%%%%%%%%%%%%%
\section{Introduction}
%%%%%%%%%%%%%%%%%%%%%%%%%%%%%%%%
A tidal disruption event (TDE) occurs when a star passes close to a black hole
(BH) and is torn apart by tidal forces. Debris from the star produces an accretion 
flare as it
falls back onto an accretion disk around the BH. Tidal disruption events and the corresponding accretion flares were first predicted by \citet{Hills75} and \citet{Rees88} respectively. Over the last decade, dozens of TDE candidates have been discovered, and the detection of various exotic transients has renewed interest in theoretical models of TDEs.  Some of the most unusual candidate events could be explained as the tidal
disruption of a white dwarf (WD) by a moderately massive black hole (MBH) \citep{Shcherbakov13}. In particular, it has been suggested that WD-TDEs could be a source of a recently discovered population of ultra long gamma ray bursts \citep{Levan14,MacLeod14,Ioka16}. 

The possibility of detecting WD-TDEs  is especially intriguing because the disruption of a WD is only visible if the mass of the BH, $M_{\rm{bh}}$, is less than about
$10^5~M_\odot$ \citep{MacLeod14}. For $M_{\rm{bh}} \gtrsim  10^5~M_\odot$, the WD is ``swallowed whole" without disruption. We expect these disruption events to be very luminous as a portion of the accretion power likely channels into a relativistic jet  \citep{Giannios11,Krolik12,Decolle12}. There are now a few cases in which  hard x-ray emission from launching a relativistic jet was observed following a flare from stellar disruption \citep{Bloom11,Burrows11,vanVelzen16}. An otherwise quiet MBH would show bright, beamed emission after the disruption of a WD. Therefore detecting (or not detecting) signals from the tidal disruptions of WDs could place constraints on the MBH  population. 

Recognizing the signal from a WD-TDE requires
a solid theoretical understanding of all pathways leading to the tidal disruption of WDs. The predicted signal from WD tidal disruption varies greatly with orbital parameters. A ``normal" TDE occurs when the pericenter distance ($r_{\rm{p}}$) between the star and the BH is of the same order, but less than the tidal radius ($r_{\rm{tide}}$), leading to a one time shredding of the star. In an extreme case, where $r_{\rm{p}}$ is much smaller than $r_{\rm{tide}}$, tidal compression of the WD could lead to a thermonuclear explosion, which would produce a distinctive signal \citep{Luminet89,Rosswog08a,Rosswog08b,Rosswog09,MacLeod16}. If, on the other hand, $r_{\rm{p}}$ is a few times larger than $r_{\rm{tide}}$, the WD could undergo repeated tidal encounters with the BH, and may experience ``gentle" tidal stripping, producing a signal that is periodic with the orbit \citep{Zalamea10, MacLeod14}. This last case is especially intriguing because if the WD is able to complete many orbits, the system may be a source of gravitational waves, detectable by a space-based interferometer \citep{Sesana08,Zalamea10,Cheng13, East14,Cheng14, MacLeod14}.

In this paper, we consider the scenario that a WD is captured into an eccentric orbit with a pericenter distance that
is too large for the WD to suffer immediate disruption or partial mass transfer. In this case, the WD continues on its
orbit relatively intact and experiences multiple passages before disruption. When the WD passes close to the BH at pericenter, the tidal force from the BH excites oscillations and waves in the WD, transferring energy and angular momentum between the WD and its orbit. The dissipation of the excited oscillations can heat up the WD and influence its structural evolution. 

A major goal of this paper is to determine the tidal energy and angular momentum transfer rates for a WD in a high-eccentricity orbit around a MBH.  To this end, we examine the tidal excitation of gravity (buoyancy) waves, in the radiative envelope of the WD. As these waves propagate toward the stellar surface, they grow in amplitude, become non-linear and damp efficiently (Section \ref{DissipationCalculations}).  In this
scenario, waves do not reflect from the WD surface, so there are no standing waves. We calculate the 
energy and angular momentum deposition near the surface of the WD by imposing outgoing boundary conditions
on the fluid perturbation equations. \citet{Fuller12b} studied this scenario for binary WDs in circular orbits. In Section \ref{EccentricOrbit}, we generalize the method to calculate tidal energy and angular momentum transfer to a WD in an eccentric orbit around a MBH using realistic WD models generated with MESA \citep{Paxton11}. These results are presented in Section \ref{Results}. Lastly, in Section \ref{Heating} we study the structural evolution of a WD experiencing tidal heating as the orbit evolves due to gravitational radiation. 

%%%%%%%%%%%%%%%%%%%%%%%%%%
\section{Basic Scalings and Timescales}
%%%%%%%%%%%%%%%%%%%%%%%%%%

Before undertaking detailed calculations, we first consider the characteristic timescales for various physical processes associated with an eccentric WD-MBH binary. The rate of angular momentum loss due to gravitational radiation is well-known \citep{Peters64}:
	\begin{equation}
		\dot{J}_{\text{grav}} = - \frac{32}{5} \frac{M^2 M_{\rm{bh}}^2 M_{\rm{t}}^{1/2}}{c^5 a^{7/2}(1-e^2)^2}\left(1+\frac{7}{8}e^2\right),
		\label{eq:JdotGrav}	
	\end{equation}
where $M$ is the mass of the WD, $M_{\rm{bh}} \gg M$ is the mass of the BH, and $M_{\rm{t}} = M_{\rm{bh}} + M \simeq M_{\rm{bh}}$ is the total mass. The semi-major axis and eccentricity of the orbit are $a$ and $e$ respectively. The orbital angular momentum is given by
	\begin{equation}
	J = \mu\sqrt{G M_{\rm{t}}a(1-e^2)},
	\label{eq:JOrb}
	\end{equation}
where $\mu$ is the reduced mass of the system. We define
	\begin{equation}
		r_{\rm{t}} \equiv R\left(\frac{M_{\rm{t}}}{M}\right)^{1/3},
		\label{eq:rtDef}
	\end{equation}
where $R$ is the WD radius, so that the tidal radius for stellar disruption is a few times $r_{\rm{t}}$. It is convenient to define the dimensionless pericenter distance,
	\begin{equation}\label{eq:etadef}
	\eta \equiv \frac{r_{\rm{p}}}{r_{\rm{t}}} =  (1-e) \left(\frac{\Omega_*}{\Omega}\right)^{2/3}
	\end{equation}
	where $\Omega = \sqrt{GM_{\rm{bh}}/a^3}$ is the orbital angular frequency and $\Omega_* = \sqrt{GM/R^3}$. Note that tidal disruption occurs when $\eta \lesssim 3$. 

	%%%%%Figure%%%%%
	\begin{figure*}
	\includegraphics[width=6 in]{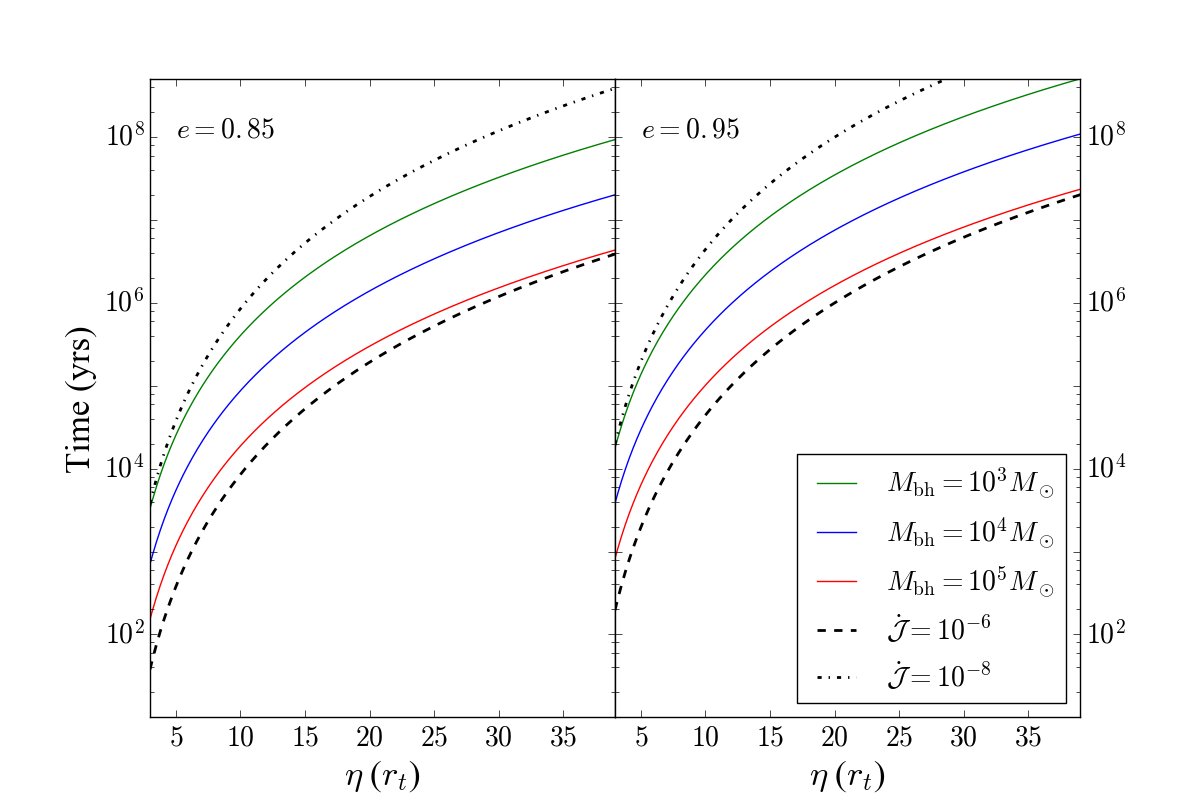}
	\caption{The orbital evolution timescales from equation (\ref{eq:tGrav}) for gravitational radiation (solid colored lines) for three different 		BH masses. The synchronization timescales (dotted lines)  are also shown for two values of $\dot{\mathcal{J}}$. Note that the 		synchronization time [see equation (\ref{eq:tSynch})] is independent of $M_{\rm{bh}}$ when plotted vs. $\eta$. The WD mass is $M = 			0.6 M_\odot$ and its radius is $R = 0.012 R_\odot$.The left and right panels display results for $e=0.85$ and $e=0.95$ 			respectively. }
	\label{fig:timeScales}
	\end{figure*}

Throughout the paper, we assume that disruption occurs outside of the Schwarzschild radius, so  $r_{\rm{t}} \gtrsim 2 G M_{\rm{t}}/c^2$ or 
	\begin{equation}
	\frac{M_{\rm{t}}}{M} \lesssim 3.2 \times 10^5 \left(\frac{R}{0.012 R_\odot}\right)^{3/2}\left(\frac{M}{0.6 M_\odot}\right)^{-3/2}.
	\end{equation}
In terms of $\eta$, the orbital evolution timescale due to gravitational radiation is (for $M_{\rm{bh}} \gg M$)
	\begin{equation}
		t_{\text{grav}} = \frac{J}{|\dot{J}_{\text{grav}}|} = \frac{5}{32}\frac{c^5 R^4 \eta^4}{G^3M^{7/3}M_{\rm{bh}}^{2/3}(1-e)^{3/2}}
		\frac{(1+e)^{5/2}}{(1 + 7e^2/8)}.
	\end{equation}
Aside from the factor of $(1-e)^{-3/2}$, $t_{\rm{grav}}$ depends rather weakly on $e$.  Around $e=0.95$, we have
	\begin{align}
		t_{\text{grav}} =& 1.01 \times 10^5~\text{yrs }\left(\frac{\eta}{10}\right)^4\left(\frac{R}{0.012 R_{\odot}}\right)^4 \nonumber\\
		&\times \left(\frac{M}{0.6 M_{\odot}}\right)^{-7/3}\left(\frac{M_{\rm{bh}}}{10^5 M_{\odot}}\right)^{-2/3}\left(\frac{1-e}{0.05}\right)^{-3/2}.
		\label{eq:tGrav}
	\end{align}
The timescale for gravitational radiation is shown as a function of $\eta$ in Fig. \ref{fig:timeScales}.

To estimate the timescale associated with tidal dissipation, we must know the rates of energy and angular momentum transfer via tides. In Sections \ref{DissipationCalculations} and \ref{EccentricOrbit}, we present our calculations of these rates. A simple parameterization of the energy and angular momentum transfer rate can be obtained as follows. For a highly eccentric orbit, tidal torque on the WD mainly occurs near pericenter, and is of order
	\begin{equation}
	T_{\rm{p}} \sim \frac{GM_{\rm{bh}}^2R^5}{r_{\rm{p}}^6} \delta_{\rm{p}},
	\end{equation}
where $\delta_{\rm{p}}$ is the tidal lag angle (of order the inverse of the tidal quality factor, $Q$). To obtain the orbit-averaged torque $\langle T \rangle$, we scale $T_{\rm{p}}$ by the ratio of $\Omega$ to the orbital angular velocity at pericenter
	\begin{equation}
	 \Omega_{\rm{p}}=\frac{\Omega}{(1-e)^{3/2}}=\left(\frac{GM_{\rm{bh}}}{a^3}\right)^{1/2}\frac{1}{(1-e)^{3/2}}. 
	\end{equation}
Thus,
 	\begin{equation}
		\langle T \rangle \sim \frac{GM_{\rm{bh}}^2R^5}{r_{\rm{p}}^6} \delta_{\rm{p}} \left(\frac{\Omega}{\Omega_{\rm{p}}}\right).
	\end{equation}
Motivated by this expression, we define the dimensionless tidal angular momentum transfer rate, $\dot{\mathcal{J}}$, via
	\begin{equation}\label{eq:dimJdef}
		\langle T \rangle \equiv \dot{J}_{\text{tide}} = \frac{G M_{\rm{bh}}^2R^5}{r_{\rm{p}}^6}(1-e)^{3/2}\dot{\mathcal{J}}.
	\end{equation}
Similarly, we define the dimensionless tidal energy transfer rate in the inertial frame, $\dot{\mathcal{E}}_{\rm{in}}$, via
	\begin{equation}\label{eq:dimEdef}
		\dot{E}_{\text{tide,in}}= \frac{G M_{\rm{bh}}^2R^5}{r_{\rm{p}}^6}\Omega \; \dot{\mathcal{E}}_{\rm{in}}.
	\end{equation}
When the WD has a finite rotation rate $\Omega_{\rm{s}}$, we will also calculate the tidal energy transfer rate in the rotating frame,
	\begin{equation}\label{eq:dimErotdef}
	\dot E_{\rm{tide,rot}} =\dot E_{\rm{tide,in}} -\Omega_{\rm{s}} \dot J_{\rm{tide}} = \frac{G M_{\rm{bh}}^2R^5}{r_{\rm{p}}^6}\Omega \; \dot{\mathcal{E}}_{\rm{rot}}.
	\end{equation}
Our calculations in Sections \ref{DissipationCalculations} and \ref{EccentricOrbit} suggest that $\dot{\mathcal{J}}$ ranges from $\lesssim10^{-8}$ for $\eta \gtrsim10$ to $10^{-6}$ for $\eta \sim$ a few. 

The timescale for orbital evolution due to dynamical tides (for $M_{\rm{bh}} \gg M$) is given by 
	\begin{equation}
		t_{\text{tide}} = \frac{J}{|\dot{J}_{\text{tide}}|} = \frac{M_{\rm{bh}}^{2/3} R^{3/2} \eta^{13/2}}{G^{1/2} M^{7/6} \dot{\mathcal{J}}}
		\frac{(1+e)^{1/2}}{(1-e)^{3/2}}. 
	\end{equation}
For an eccentricity of $e=0.95$, we have
	\begin{align}
		t_{\text{tide}} \simeq & 1.03 \times 10^{11} ~\text{yrs} \left(\frac{\eta}{10}\right)^{13/2}\left(\frac{R}{0.012 R_{\odot}}\right)^{3/2}\left(\frac{M}{0.6 M_{\odot}}\right)^{-7/6} \nonumber\\
		&\times \left(\frac{M_{\rm{bh}}}{10^5 M_{\odot}}\right)^{2/3} \left(\frac{\dot{\mathcal{J}}}{10^{-6}}\right)^{-1}\left(\frac{1-e}{0.05}\right)^{-3/2}.
		\label{eq:tTides}
	\end{align}
Comparing $t_{\text{grav}}$ and $t_{\text{tide}}$ shows that gravitational radiation dominates the orbital evolution. For an intermediate to massive BH, the gravitational radiation timescale is always much shorter than the tidal dissipation timescale, even for small pericenter distances of $\eta \sim 3$.

The tidal torque can also affect the rotation rate of the WD, driving it toward pseudo-synchronization, such that the WD rotation rate $\Omega_{\rm{s}}$ approaches a value comparable to the orbital frequency at pericenter  $\Omega_{\rm{p}}$. The tidal synchronization timescale can be estimated by
	\begin{equation}
		t_{\text{synch}} =  \frac{I \Omega_{\rm{p}}}{|\dot{J}_{\text{tide}}|} = \left(\frac{G M}{R^3}\right)^{-1/2}\frac{k \eta^{9/2}}{(1-e)^{3/2}\dot{\mathcal{J}}},
	\end{equation}
	where $I=kMR^2$ is the moment of inertia of the WD. Note that the synchronization timescale is independent of the BH mass when written in terms of $\eta$. For $k = 0.17$, we have
	\begin{align}
		t_{\text{synch}} =& 4.1 \times 10^4~\text{yrs} \left(\frac{\eta}{10}\right)^{9/2}\left(\frac{R}{0.012 R_{\odot}}\right)^{3/2} \nonumber\\
		&\times \left(\frac{M}{0.6 M_{\odot}}\right)^{-1/2} \left(\frac{\dot{\mathcal{J}}}{10^{-6}}\right)^{-1}\left(\frac{1-e}{0.05}\right)^{-3/2}.
		\label{eq:tSynch}
	\end{align}
For $\dot{\mathcal{J}} = 10^{-6}$, the synchronization timescale is shorter than the gravitational radiation timescale. In this case, the WD would be rotating at a rate $\Omega_{\rm{s}}\sim\Omega_{\rm{p}}$ before tidal disruption, and rotation must be included in calculations of tidal dissipation. However, if $\dot{\mathcal{J}}$ is smaller by even a factor of 100, the WD will not synchronize before disruption (see Fig. \ref{fig:timeScales}). Results in Section \ref{Results} show that, at large separations, $\dot{\mathcal{J}}$ can be as small as $10^{-10}$, implying that $t_{\rm{synch}} \gg t_{\rm{grav}}$. In this situation, the WD will rotate with its rotation rate at capture until its eventual disruption. For a captured isolated WD, we expect $\Omega_{\rm{s}} \ll \Omega_{\rm{p}}$, thus rotation has a small effect on dynamical tides. A WD that originated in a binary may still be rotating rapidly after capture, in which case the rotational effect can be significant.

%%%%%%%%%%%%%%%%%%%%%%%%%%%%%%%%%%%%%	
\section{WD Model and Physics of Tidal Dissipation}\label{DissipationCalculations}
%%%%%%%%%%%%%%%%%%%%%%%%%%%%%%%%%%%%%

%Changed for single column
	%%%%%Figure%%%%%
	\begin{figure}
		\includegraphics[width=\columnwidth]{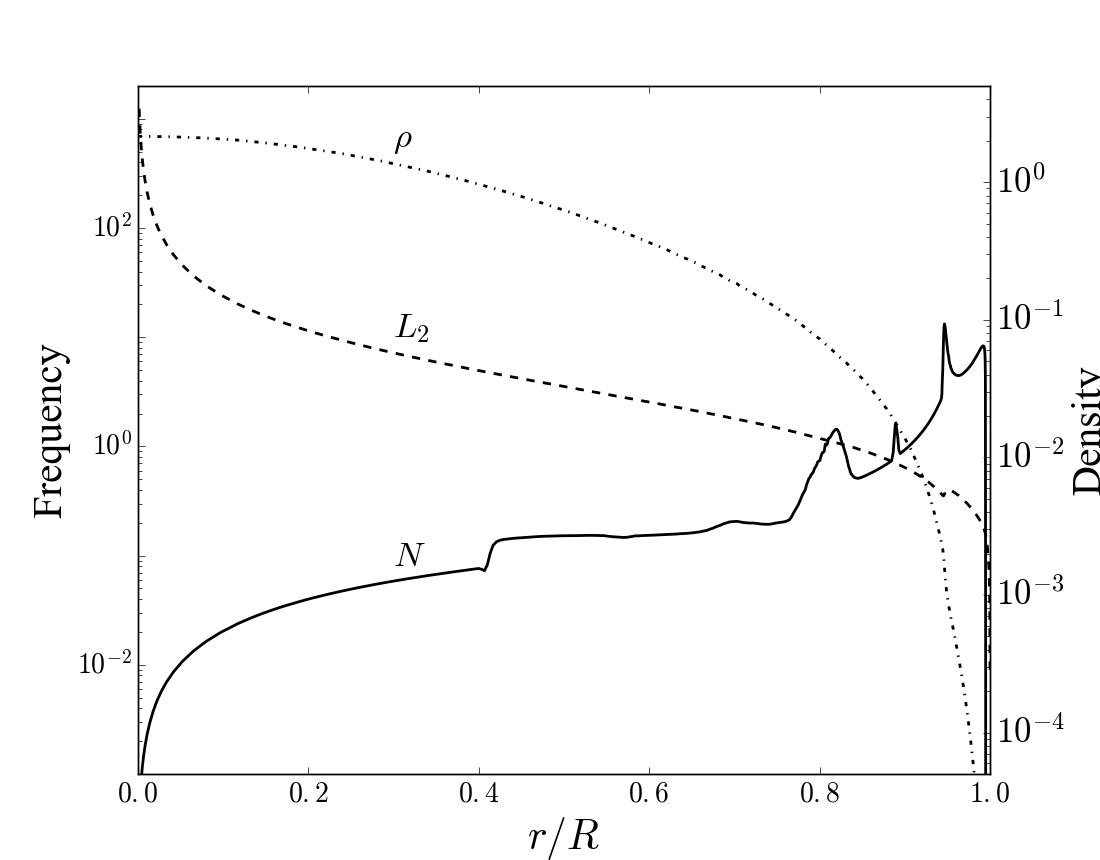}
		\includegraphics[width=\columnwidth]{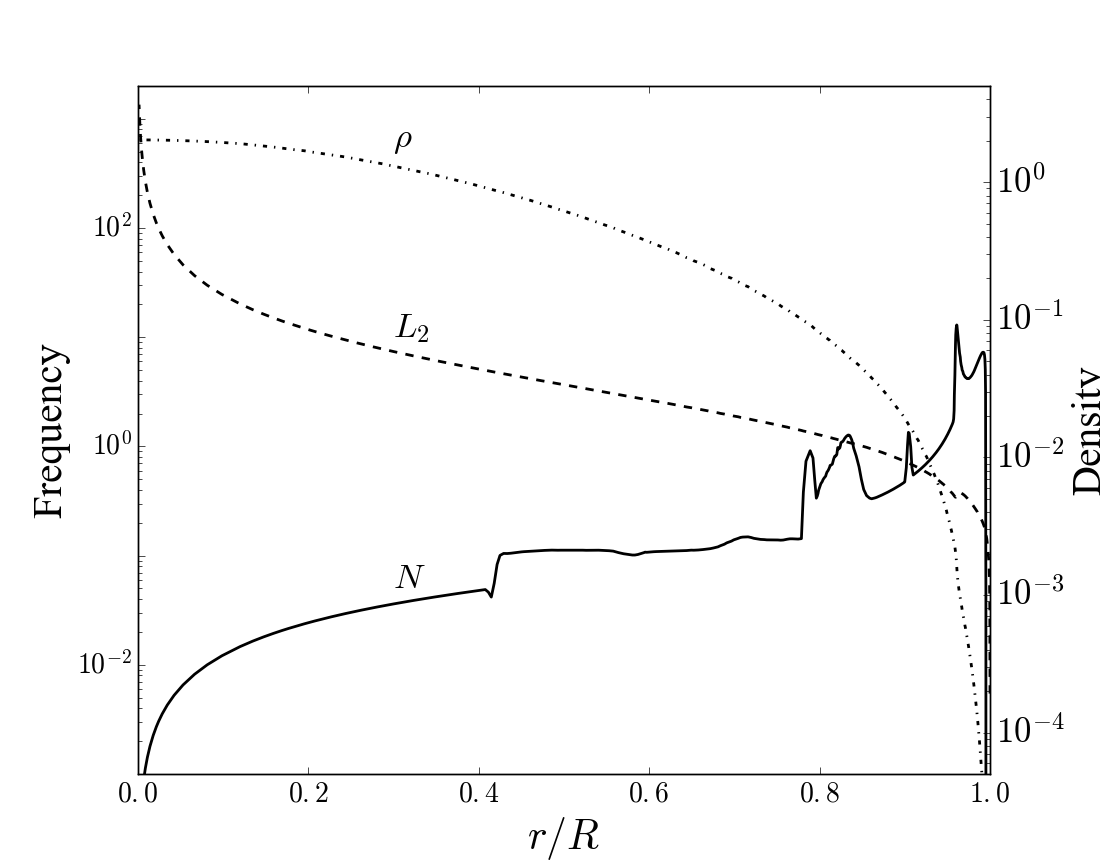}
		\caption{Propagation diagrams for MESA-generated WD models with mass $M = 0.61 M_\odot$ and 
		radius $R = 0.012 R_\odot$. The WD model in the upper plot has $T_{\textrm{eff}} = 9,000$~K. The model in the lower plot has $T_{\textrm{eff}} = 5,200$~K. The diagram includes the Br\"unt-V\"ais\"al\"a freqency $N$ (solid line), the 				Lamb frequency $L_2$ (dashed line) and the density $\rho$ (dot-dashed line).  Frequencies 				are in units of $(GM/R^3)^{1/2}$, and densities in units of $(M/R^3)$. The peaks in $N$ generally 				correspond to compositional transitions from O to C, C to He, and He to H.}			
		\label{fig:PropDiagramCool}
\end{figure}
	
	Tidal dissipation in WDs arises from the excitation of gravity waves in the deeper envelope of the WD and their dissipation in the outer envelope \citep{Fuller12c,Fuller13}. These processes depend on the detailed properties of the WD, in particular, properties of the envelope. In this paper, we consider two CO WD models consisting of a CO core with a He-H envelope. The models are constructed using the MESA stellar evolution code  \citep{Paxton11}. Both WD models have mass $M = 0.61 M_\odot$. One model has effective temperature $T_{\textrm{eff}} = 9,000$~K. The other was allowed to evolve for a longer time and is therefore cooler with $T_{\textrm{eff}} = 5,200$~K. 

	Fig. \ref{fig:PropDiagramCool} displays the profiles for density $\rho$, Lamb (acoustic cutoff) frequency $L_2$ and the  Br\"unt-V\"ais\"al\"a frequency $N$ for the two WD models. The Lamb frequency $L_l$ is given by
	\begin{equation}
	L_l^2 = \frac{l(l+1) a_{\rm{s}}^2}{r^2},
	\label{eq:Lamb}
	\end{equation}
where $a_{\rm{s}}$ is the sound speed. The  Br\"unt-V\"ais\"al\"a frequency is given by 
	\begin{equation}
	N^2 = - g\left(\frac{1}{\rho} \frac{d\rho}{dr} + \frac{g}{a_{\rm{s}}^2}\right),
	\label{eq:BVfreq}
	\end{equation}
where $g$ is gravitational acceleration. Gravity waves propagate in regions where $\omega<N$ and $\omega<L_l$, and become evanescent elsewhere. We see from Fig. \ref{fig:PropDiagramCool} that the propagation diagrams for both WD models exhibit peaks in the $N$-profile that are associated with composition changes from CO to He and He to H. These peaks occur in similar locations in the two models but have somewhat different structures.

\begin{figure*}
		\includegraphics[width=6in]{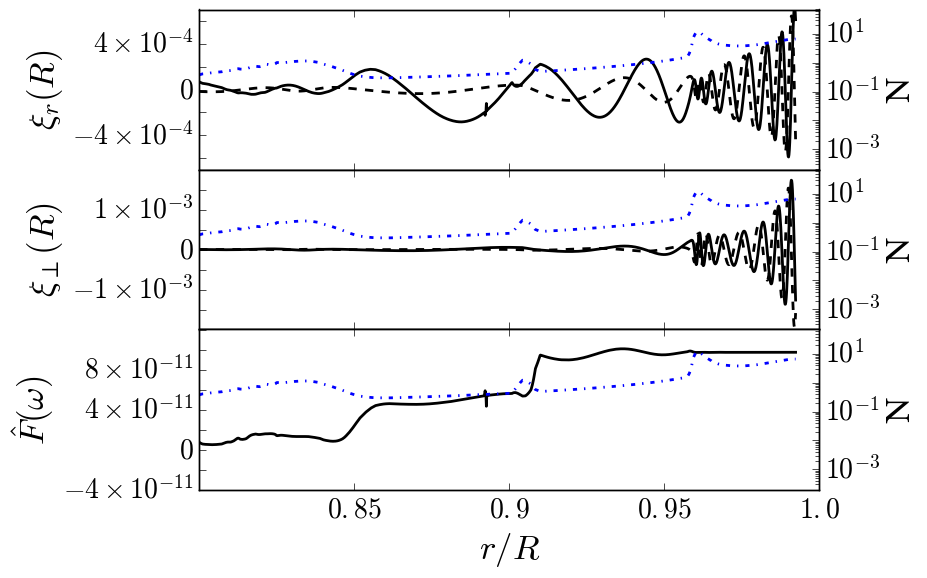}
		\caption{Dynamical tide in a realistic MESA-generated WD model produced by a companion in a circular orbit. The WD has 		mass $M = 0.61 M_\odot$ and radius $R = 0.012 M_\odot$ with $T_{\textrm{eff}} = 5,200$~K. Tides are excited by a 			companion with mass $M_{\rm{bh}}=10^5 M_\odot$ at the frequency $\omega=0.01 (GM/R^{3})^{1/2}$. The real parts of 			the radial and transverse Lagrangian displacements, $\xi_r$ and $\xi_\perp$, are shown with solid black lines, and imaginary 			parts are shown with dotted black lines. The Br\"unt-V\"ais\"al\"a frequency in units of $(GM/R^{3})^{1/2}$ is overlaid in blue. 		Peaks correspond to compositional transitions. The dimensionless tidal torque $\hat{F}(\omega)$ begins to rise at the C/He 			transition and settles to a constant value near the stellar surface. }
		\label{fig:WaveForm}
	\end{figure*}

	\citet{Fuller12b} studied dynamical tides in WD binaries. They showed that the binary companion excites a continuous train of outgoing gravity waves, primarily around the CO/He transition region. As the waves propagate toward the stellar surface, their amplitudes grow with decreasing density. Eventually, the waves dissipate through a combination of non-linear processes, radiative damping, and absorption at a critical layer\citep{Zahn75,Goldreich89,Fuller13}. For a circular orbit, the tidal potential acting on the WD can be written as 
	\begin{equation}
		U(\textbf{r},t) = U(r)[Y_{22}(\theta,\phi)e^{-i\omega t} + c.c.],
		\label{eq:quadPotential}
	\end{equation}
	with
	\begin{equation}
		U(r) = -\frac{GM_{\rm{bh}}W_{22}r^2}{a^3},
	\end{equation}
	where $W_{22} = (3\pi/10)^{1/2}$ and $\omega = 2 \Omega$ is the tidal forcing frequency for a non-rotating WD ($\Omega_{\rm{s}}=0$). Using the method developed in \citet{Fuller12b} (similar in nature to the treatment of gravity wave damping in the center of solar-type stars from \citet{Goodman98}, \citet{Ogilvie07}, \citet{Barker10}, and \citet{Barker11}) we calculate the amplitude of outgoing gravity waves excited by the potential in equation (\ref{eq:quadPotential}) as a function of $\omega$ for the two WD models depicted in Fig. \ref{fig:PropDiagramCool}. The angular momentum flux carried by the waves is 
	\begin{equation}
	\dot{J}(r) = 2 m  \omega^2 \rho r^3 \textrm{Re}(i \xi_r^* \xi_\perp),
	\end{equation}
where $m$ is the azimuthal wave number ($m=2$ for circular binaries), and $\xi_r(r)$ and $\xi_\perp(r)$ are the radial and transverse Lagrangian displacements of the wave. Fig. \ref{fig:WaveForm} shows an example of the numerical result for $\omega = 0.01~(GM/R^3)^{1/2}$. We see that $\dot{J}(r)$ oscillates around zero in the WD interior but jumps to a constant value with the excitation of outgoing waves at compositional changes. The constant value of $\dot{J}(r)$ evaluated in the outermost region corresponds to the tidal torque acting on the WD. 
	%%%%%Figure%%%%%

\begin{figure*}
		\includegraphics[width=3.4in]{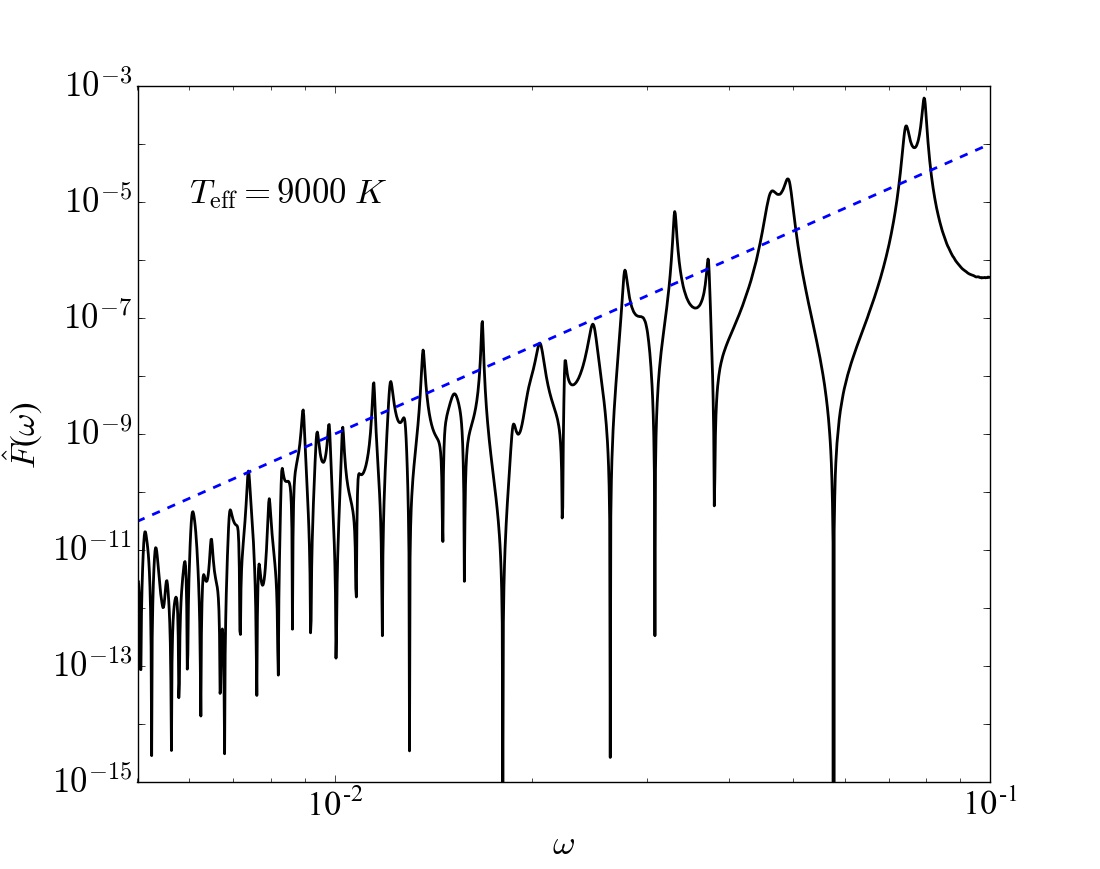}
		\includegraphics[width=3.4in]{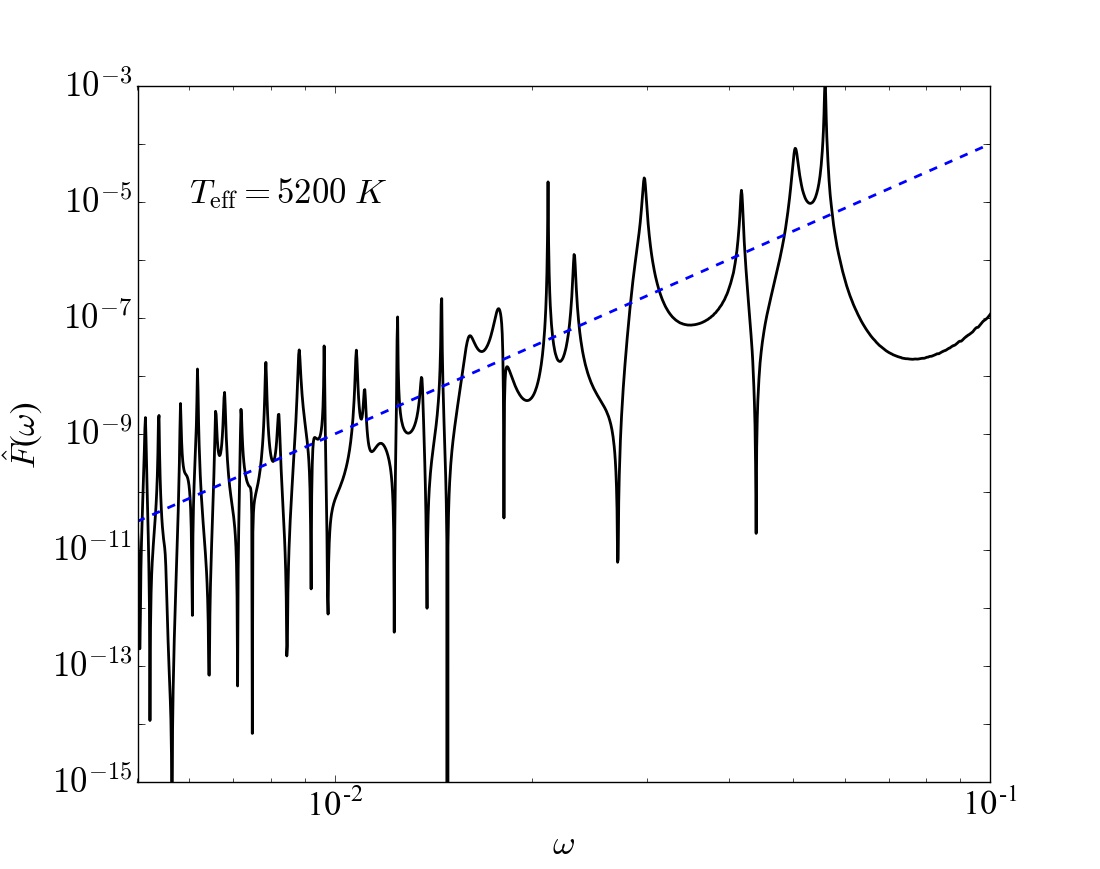}
		\caption{Dimensionless tidal torque $\hat{F}(\omega)$ as a function of the tidal forcing frequency $\omega$ in units of $(GM/R^3)^{1/2}$ (solid lines). The left 			panel is for the WD model with $M = 0.61 M_\odot$ ,  $R = 0.012 M_\odot$, and $T_{\text{eff}} =9,000$~K. The blue dotted 		line shows $10~\omega^5$, which roughly follows the maximum of $\hat{F}(\omega)$. The right panel is for a similar WD model 	with effective temperature $T_{\text{eff}} =5,200$~K. The blue dotted line shows $150~ \omega^5$.}
		\label{fig:FhatCool}
	\end{figure*}

\citet{Fuller12b} introduce a dimensionless function $\hat{F}(\omega)$ to characterize how the tidal torque depends on the tidal forcing frequency:
	\begin{equation}
	\dot{J} = G\left(\frac{M_{\rm{bh}}}{a^3}\right)^2 R^5\frac{|m|}{2}\hat{F}(\omega).
	\end{equation}
Similarly, the tidal energy transfer  rate is given by 
	\begin{equation}
	\dot{E} = G\left(\frac{M_{\rm{bh}}}{a^3}\right)^2 R^5 \Omega \hat{F}(\omega).\label{eq:EdotCirc}
	\end{equation}
Fig. \ref{fig:FhatCool} shows our numerical results for $\hat{F}(\omega)$ for the two WD models. The $\omega$ dependence of $\hat{F}(\omega)$ can be roughly understood with a simplified WD model with two regions. The outer region has a much larger Br\"unt-V\"ais\"al\"a frequency than the inner region. The inner region is like a resonance cavity. While both ingoing and outgoing waves can propagate in the inner region, the outer region has only outgoing waves. In this model, a maximum tidal torque scales as $\hat{F}_{\text{max}}(\omega)\propto \omega^5$. Our numerical results roughly agree with this scaling.
   	
%Changed for single column
	%%%%%Figure%%%%%

\section{Tidal Dissipation in an Eccentric Orbit}\label{EccentricOrbit}

	We extend the method of the previous section to study dyanmical tides in a WD in an eccentric orbit around a MBH. In this case, the quadrulpolar tidal potential can be written as
	\begin{equation}
	U = \sum_m U_{2m} (\textbf{r}_i, t),
	\end{equation}
	with
	\begin{align}
	U_{2m}(\textbf{r}_i,t) = - \frac{G M_{\rm{bh}} W_{2m} r_i^2}{D(t)^{3}} e^{-i m \Phi(t)} Y_{2m}(\theta, \phi_i),
	\end{align}
where $D(t)$ is the binary separation, $\Phi$ is the orbital phase (true anomaly), $W_{2\pm 2} = \sqrt{3\pi/10}, W_{20} = \sqrt{\pi/5}$ and $W_{2\pm1} = 0$. The vector $\textbf{r}_i = (r, \theta, \phi_i = \phi + \Omega_{\rm{s}} t)$, is the position relative to the WD center, with the azimuthal angle $\phi_i$ measured in the inertial frame ($\Omega_{\rm{s}}$ is the rotation rate of the WD, and $\phi$ is measured in the rotating frame). Each component of the potential can be decomposed into an infinite sum over forcing frequencies.
	\begin{align}
	U_{2m}=-\frac{GM_{\rm{bh}} W_{2m}r_i^2}{a^3} Y_{2m}(\theta,\phi_i) \sum_{N=-\infty}^\infty F_{Nm} e^{-iN\Omega t}, 
	\end{align}
where $F_{Nm}$ is the Hansen coefficient \citep{Murray00}
	\begin{align}
	F_{Nm} = \frac{1}{\pi} \int_{-\pi}^\pi  d\Psi \frac{\cos [ N(\Psi - e \sin \Psi) - m \Phi(t)]}{(1 - e\cos \Psi)},
	\label{eq:HansenCoeff}
	\end{align}
with $\Psi$ the eccentric anomaly. 

The total energy and angular momentum transfer rates can be expressed as a sum of the responses due to
each frequency term in the external potential.

First, consider the time-varying $m=0$ terms in the external potential,
	\begin{align}
	U_{20} = -\frac{GM_{\rm{bh}} W_{20}r^2}{a^3}\sum_{N=1}^{\infty} F_{N0} [Y_{20}(\theta,\phi_i) e^{-i N\Omega t}+ c.c.]\label{eq:m0potential}.
	\end{align}
	We have taken advantage of the fact that $F_{N0} = F_{-N0}$. Each term in equation (\ref{eq:m0potential}) has the same form as equation (\ref{eq:quadPotential}), and contributes to the energy transfer rate as in equation (\ref{eq:EdotCirc}):
	\begin{align}\dot{E}_{m=0} = \frac{G M_{\rm{bh}}^2 R^5}{a^6}\left(\frac{W_{20}}{W_{22}}\right)^2\sum_{N=1}^{\infty} N \Omega F_{N0}^2 \hat{F}(\omega=|N\Omega|). 
	\end{align}
	The factor $(W_{20}/W_{22})^2$ arises because $\hat{F}(\omega)$ includes $(W_{22})^2$. The $m=0$ potential does not contribute to angular momentum transfer because it is axisymmetric. 

	We can determine the energy transfer rate due to $m=\pm2$ components of the potential using similar procedures. In the rotating frame of the WD,
	\begin{align}
	U_{22} =& - \frac{GM_{\rm{bh}} W_{22} r^2}{a^3} \sum_{N=1}^{\infty}\left[F_{N2} Y_{22} (\theta, \phi)e^{-i(N\Omega  - 2 \Omega_{\rm{s}} )t } \right.\nonumber\\ 
		&\left.+ F_{-N2} Y_{22}(\theta, \phi) e^{i (N\Omega  + 2 \Omega_{\rm{s}} )t}\right],\\
	U_{2-2} =& - \frac{GM_{\rm{bh}} W_{22} r^2}{a^3} \sum_{N=1}^{\infty}\left[F_{-N2} Y_{2-2} (\theta, \phi)e^{-i(N\Omega  + 2 \Omega_{\rm{s}} )t }\right.\nonumber\\ 
	&\left.+ F_{N2} Y_{2-2}(\theta, \phi) e^{i (N\Omega  - 2 \Omega_{\rm{s}} )t}\right].
	\end{align}
	Using symmetries of the spherical harmonics, the sum of the $m=2$ and $m=-2$ potentials reduces to 
	\begin{align}
	 U_{22} + U_{2-2} =& - \frac{GM_{\rm{bh}} W_{22} r^2}{a^3}\nonumber \\
			& \times \sum_{N=-\infty}^{\infty}\left[ F_{N2} Y_{22}(\theta,\phi) e^{-i(N\Omega - 2 \Omega_{\rm{s}} )t}+ c.c. \right] \label{eq:U2m2}.
	\end{align}
As with the $m=0$ terms of the potential, we can translate each term into an energy transfer rate of the same form as equation (\ref{eq:EdotCirc}). However, equation (\ref{eq:U2m2}) differs from equation (\ref{eq:m0potential}) in that it includes both negative and positive $\omega$, so we must interpret the physical meaning of the contributions from the negative frequency terms. In a frame that rotates with the WD, positive $\omega$ corresponds to the BH orbiting the WD in a counterclockwise direction. Negative $\omega$ corresponds to a clockwise orbit. Therefore, changing the sign of $\omega$ reverses the sign of the angular momentum transfer rate, but does not affect the sign of the energy transfer rate (in the rotating frame). 

	Combining contributions from $m=0,-2$ and $2$ gives the following expressions for angular momentum and energy transfer 
	 \begin{align}
	\dot{J}_{\rm{tide}} =& T_0 \sum_{-\infty}^{\infty} F_{N2}^2 \sgn(N\Omega - 2 \Omega_{\rm{s}})\hat{F}(\omega =  |N\Omega - 2 \Omega_{\rm{s}}|),\label{eq:Jrate}\\
	\dot{E}_{\textrm{tide,in}}=& T_0 \left[\left(\frac{W_{20}}{W_{22}}\right)^2\sum_{N=1}^{\infty}N \Omega F_{N0}^2 \hat{F}(\omega = |N\Omega|)\right. \nonumber \\ & \left.+  \frac{1}{2}\sum_{-\infty}^{\infty} N \Omega F_{N2}^2 \sgn(N\Omega - 2 \Omega_{\rm{s}})\hat{F}(\omega =  |N\Omega - 2 \Omega_{\rm{s}}|) \right] \label{eq:Einrate},\\
	\dot{E}_{\text{tide,rot}}=& T_0 \left[\left(\frac{W_{20}}{W_{22}}\right)^2\sum_{N=1}^{\infty}N \Omega F_{N0}^2 \hat{F}(\omega = |N\Omega|)\right. \nonumber \\ & \left.+  \frac{1}{2}\sum_{-\infty}^{\infty}  F_{N2}^2 |N\Omega - 2 \Omega_{\rm{s}}|\hat{F}(\omega = |N\Omega - 2 \Omega_{\rm{s}}|) \right]\label{eq:Erotrate},
	\end{align}
with 
	\begin{equation}
	T_0 \equiv \frac{G M_{\rm{bh}}^2 R^5}{a^6}.
	\end{equation}
Note that the energy transfer rate in the inertial frame ($\dot{E}_{\rm{tide,in}}$) is related to that in the rotating frame through equation (\ref{eq:dimErotdef}).

We can determine the dimensionless quantities $\dot{\mathcal{J}}$, $\dot{\mathcal{E}}_{\rm{in}}$, and $\dot{\mathcal{E}}_{\rm{rot}}$ from equations (\ref{eq:dimJdef})-(\ref{eq:dimErotdef}).
	\begin{align}
	\dot{\mathcal{J}} =& (1-e)^{9/2} \sum_{-\infty}^{\infty} F_{N2}^2 \sgn(N\Omega - 2 \Omega_{\rm{s}}) \hat{F}(\omega =  |N\Omega - 2 \Omega_{\rm{s}}|)\label{eq:dimJ}.\\
	\dot{\mathcal{E}}_{\textrm{in}}=&  (1-e)^6 \left[\left(\frac{W_{20}}{W_{22}}\right)^2\sum_{N=1}^{\infty}N F_{N0}^2 \hat{F}(\omega = |N\Omega|)\right. \nonumber \\ & \left.+  \frac{1}{2}\sum_{-\infty}^{\infty} N  F_{N2}^2 \sgn(N\Omega - 2 \Omega_{\rm{s}})\hat{F}(\omega = |N\Omega - 2 \Omega_{\rm{s}}|) \right], \label{eq:dimE}\\
	\dot{\mathcal{E}}_{\text{rot}}=&  (1-e)^6 \left[\left(\frac{W_{20}}{W_{22}}\right)^2\sum_{N=1}^{\infty}N F_{N0}^2 \hat{F}(\omega = |N\Omega|)\right. \nonumber \\ & \left.+  \frac{1}{2}\sum_{-\infty}^{\infty}  \hat{F}_{N2}^2 \left|N - 2 \frac{\Omega_{\rm{s}}}{ \Omega}\right|\hat{F}(\omega = |N\Omega - 2 \Omega_{\rm{s}}|) \right].\label{eq:dimErot}
	\end{align}

	When calculating transfer rates, we used two conditions to truncate the sums in equations (\ref{eq:dimJ})-(\ref{eq:dimErot}). The first is the physical condition that gravity waves with frequencies larger than $\omega \sim 0.1~(GM/R^{3})^{1/2}$ cannot propagate near the surface. This can be seen in the propagation diagrams for our two numerical WD models (see Fig. \ref{fig:PropDiagramCool}). Larger frequencies do not satisfy the conditions  $\omega < N$ and $\omega < L_2$ in the outer envelope of the WD.
Additionally, the WKB approximation used to calculate $\hat{F}(\omega)$ becomes less reliable for $\omega \sim 0.1\gtrsim(GM/R^{3})^{1/2}$. We therefore assume that terms in the tidal potential with these frequencies do not contribute to the tidal transfer rates. Physically, this assumption is reasonable because high-frequency gravity waves either reflect from the stellar surface or do not suffer non-linear breaking, reducing their ability to contribute to tidal energy dissipation. Nevertheless, for small pericenter separations, this cut-off can significantly limit the number of allowed terms in equations (\ref{eq:dimJ})-(\ref{eq:dimErot}), perhaps limiting the reliability of the calculated transfer rates.

	The second truncation condition accounts for the fact that the Hansen coefficients fall off for large values of $N$. Terms with $N \gtrsim 100\;\Omega_{\rm{p}}/\Omega$ do not contribute significantly to the transfer rates and  are not included in the sums. Even with this cutoff, some values of $F_{Nm}$ for large $e$ and $N$ cannot be efficiently calculated by numerically integrating equation (\ref{eq:HansenCoeff}). This is handled by approximating $F_{Nm}$ for large $N$ as a function of the form $F_m(N) = \alpha N^\beta \rm{exp}(-\gamma N)$ where $\alpha, \beta$, and $\gamma$ are determined by fitting $F_m(N)$ for smaller $N$. The approximation is described in fuller detail in Appendix \ref{sec:ApA}.
	
\begin{figure*}
		\includegraphics[width= 5 in]{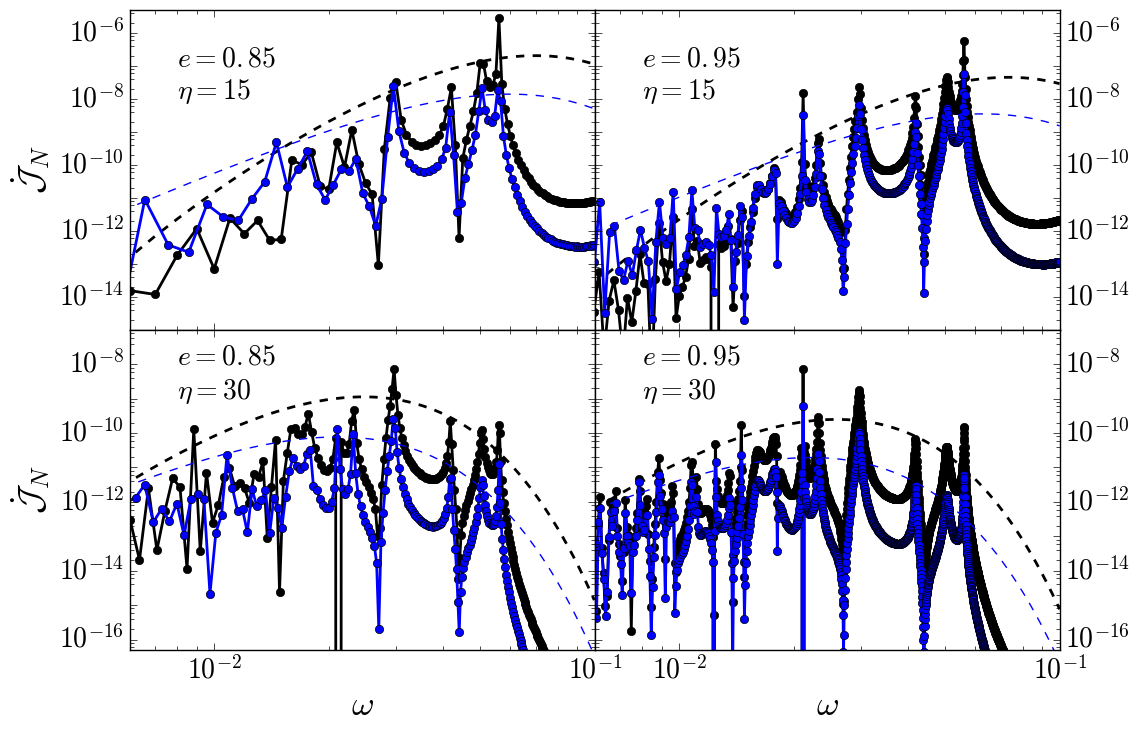}
		\caption{The individual terms $\dot{\mathcal{J}}_N$ that sum to $\dot{\mathcal{J}}$ as a function of frequency 			$\omega$ in units of $(GM/R^3)^{1/2}$ for the cooler WD model. Frequency is related to $N$ by equation (\ref{eq:omegaN}). The dotted lines show the same 			calculations using $\hat{F}(\omega) = 150 \;\omega^5$.  The black lines show results for $\Omega_{\rm{s}}=0$. The blue 		lines show results for $\Omega_{\rm{s}} = \Omega_{\rm{p}}$,  and roughly correspond to a system with a pseudo-synchronously 			rotating WD. For the bottom row with $\eta=30$, the largest peak corresponds to a lower $\omega$ than in the upper panels. Negative frequency terms are not included because their contributions are relatively small for large $\eta$.}
		\label{fig:FreqContLargeEta}
	\end{figure*}

\begin{figure*}
		\includegraphics[width= 5 in]{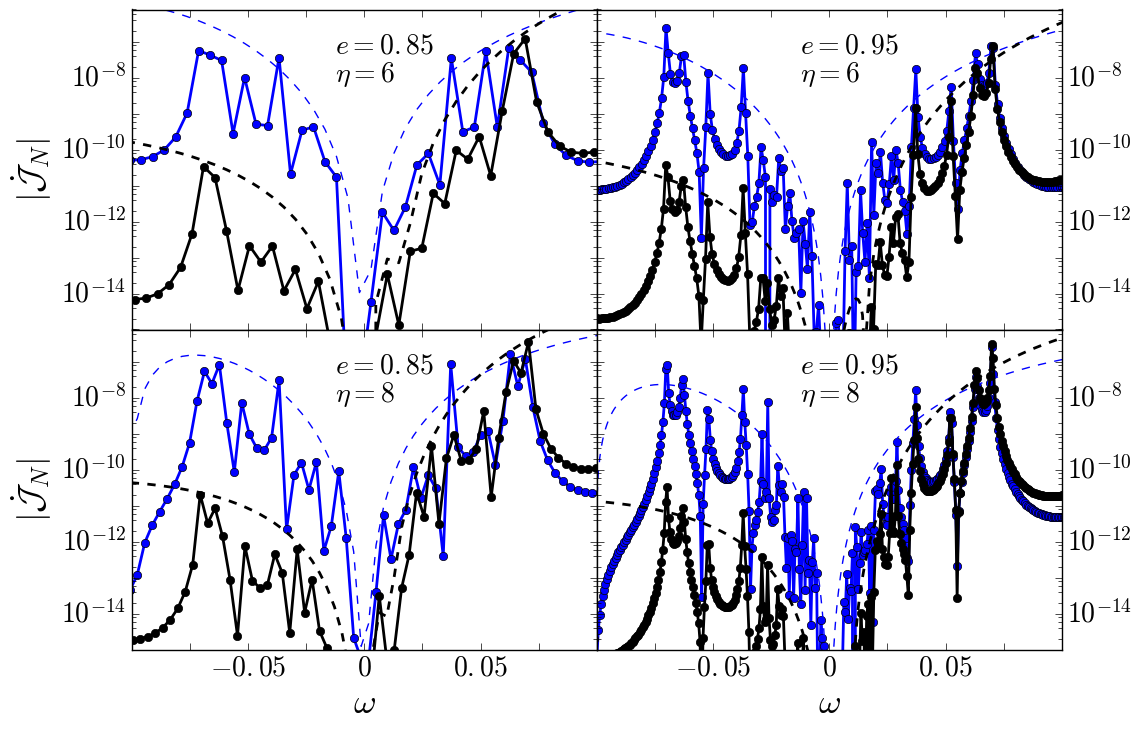}
		\caption{ Similar to Fig. \ref{fig:FreqContLargeEta}, but for smaller values of $\eta$ and with negative frequencies included. Negative frequency terms are most important for 			systems with rotating WDs and smaller values of $\eta$. Because calculations of $\hat{F}(\omega)$ do not include the Coriolis 			force, terms with $\Omega_{\rm{s}}>|\omega|$ may be less reliable. Unfortunately, this stringently limits the number of reliable 			terms for small $\eta$ and large $\Omega_{\rm{s}}$. In the top panels, $\Omega_s=0.068~(GM/R^3)^{1/2}$, and none of the plotted terms 			meet the above criterion. In the bottom panels, $\Omega_s=0.044~(GM/R^3)^{1/2}$.}
		\label{fig:FreqContSmallEta}
	\end{figure*}

	To better understand which frequencies dominate the tidal transfer rates, we can examine individual components of the sum
	\begin{equation}
		\dot{\mathcal{J}} = \sum_{N} \dot{\mathcal{J}}_N.
	\end{equation}
Each value of $N$ corresponds to an $\omega$ via the relationship $\omega = N\Omega - 2 \Omega_{\rm{s}}$. In units of $(GM/R^3)^{1/2}$, 
	\begin{equation}\label{eq:omegaN}
		\omega = \left(\frac{1}{\eta}\right)^{3/2}\left[N(1-e)^{3/2}-2\frac{\Omega_{\rm{s}}}{\Omega_{\rm{p}}}\right].
	\end{equation}
Figs. \ref{fig:FreqContLargeEta} and \ref{fig:FreqContSmallEta} show examples of the individual terms that contribute to $\dot{\mathcal{J}}$ for systems with different $e$ and $\eta$. These figures display a few trends that are discussed further in Section \ref{Results}. First, for $\eta \gtrsim 10$, increasing $\eta$ lowers both the transfer rates and the dominant frequency in the sums (the frequency corresponding to the largest $\dot{\mathcal{J}}_N$). Second, increasing $e$ and $\eta$ significantly increases the number of terms in the sums. This tends to smooth out the erratic dependence of $\hat{F}(\omega)$ on $\omega$. Third, increasing $\Omega_{\rm{s}}$ tends to lower the transfer rates.  Lastly, for $\eta \lesssim 10$ and $\Omega_s \sim \Omega_{\rm{p}}$, negative frequencies can contribute significantly to the sums. 
	
	There are a few limitations to the calculations of $\dot{\mathcal{J}}$, $\dot{\mathcal{E}}_{\rm{in}}$, and $\dot{\mathcal{E}}_{\rm{rot}}$ as presented in equations (\ref{eq:dimJ})-(\ref{eq:dimErot}). First, the Coriolis force is not included in the computation of $\hat{F}(\omega)$. The effects of rotation are handled solely by modifying the external potential. For the scenarios considered in this paper we do not expect a full treatment of rotation to significantly alter our results. \citet{Fuller14} demonstrated that increasing $\Omega_s/\omega$ has little effect on $\hat{F}(\omega)$ for a subsynchronous CO WD in which prograde g-waves contribute most to the tidal torque. However, the most dependable terms in the tidal transfer rate sums satisfy $\omega \gtrsim \Omega_s$ or
	\begin{equation}
	N(1 - e)^{3/2} \gtrsim 3.
	\end{equation} 
For a system with smaller $\eta$, few (if any) of the terms that satisfy this condition will also satisfy $\omega<0.1~(GM/R^{3})^{1/2}$. Furthermore, systems with small  $\eta$ are more likely to have large $\Omega_s$ as they have shorter synchronization timescales. The calculated transfer rates may be less reliable for small $\eta$ and large $\Omega_s$. Calculations for $\eta\gtrsim 10$ are unaffected by this problem. 

	Another limitation is that (as noted above) the WKB approximation used to compute $\hat{F}(\omega)$ breaks down for $\omega \gtrsim 0.1~(GM/R^{3})^{1/2}$. Once again, only smaller values of $\eta$ are affected by this limitation. Recall that  $\hat{F}(\omega)$ tends to increase steeply with $\omega$. Unless the Hansen coefficients, $F_{Nm}$, that correspond to large $N\Omega$ are very small, the highest frequency terms dominate the tidal transfer rate sums. The Hansen coefficients fall off rapidly with increasing $N$, so the lower $\Omega$, the lower the dominant forcing frequency and the more reliable the calculated transfer rates. In general, results from systems with large $\eta$ and small $\Omega_{\rm{s}}$ are the most reliable.

	%%%%%Figure%%%%%

	%%%%%Figure%%%%%

%%%%%%%%%%%%%%%%%%%%%%%%%%%%%%%%%%%%%%%
\section{Results for Tidal Energy and Angular Momentum Transfer Rates}\label{Results}
%%%%%%%%%%%%%%%%%%%%%%%%%%%%%%%%%%%%%%%
	
	Using the procedure and assumptions described in Section \ref{EccentricOrbit}, we calculate the dimensionless tidal transfer rates $\dot{\mathcal{E}}_{\text{in}}$, $\dot{\mathcal{E}}_{\text{rot}}$ and $\dot{\mathcal{J}}$ for both WD models for a variety of $\eta$, $e$ and $\Omega_{\rm{s}}$. To discuss the results, we will consider two different regimes --- the far regime, where $\eta \gtrsim 10$ (the pericenter distance is more than 10 times the tidal radius), and the close regime, where $\eta \lesssim 10$. 
	%%%%%Figure%%%%%
	\begin{figure*}
		\includegraphics[width = 7 in]{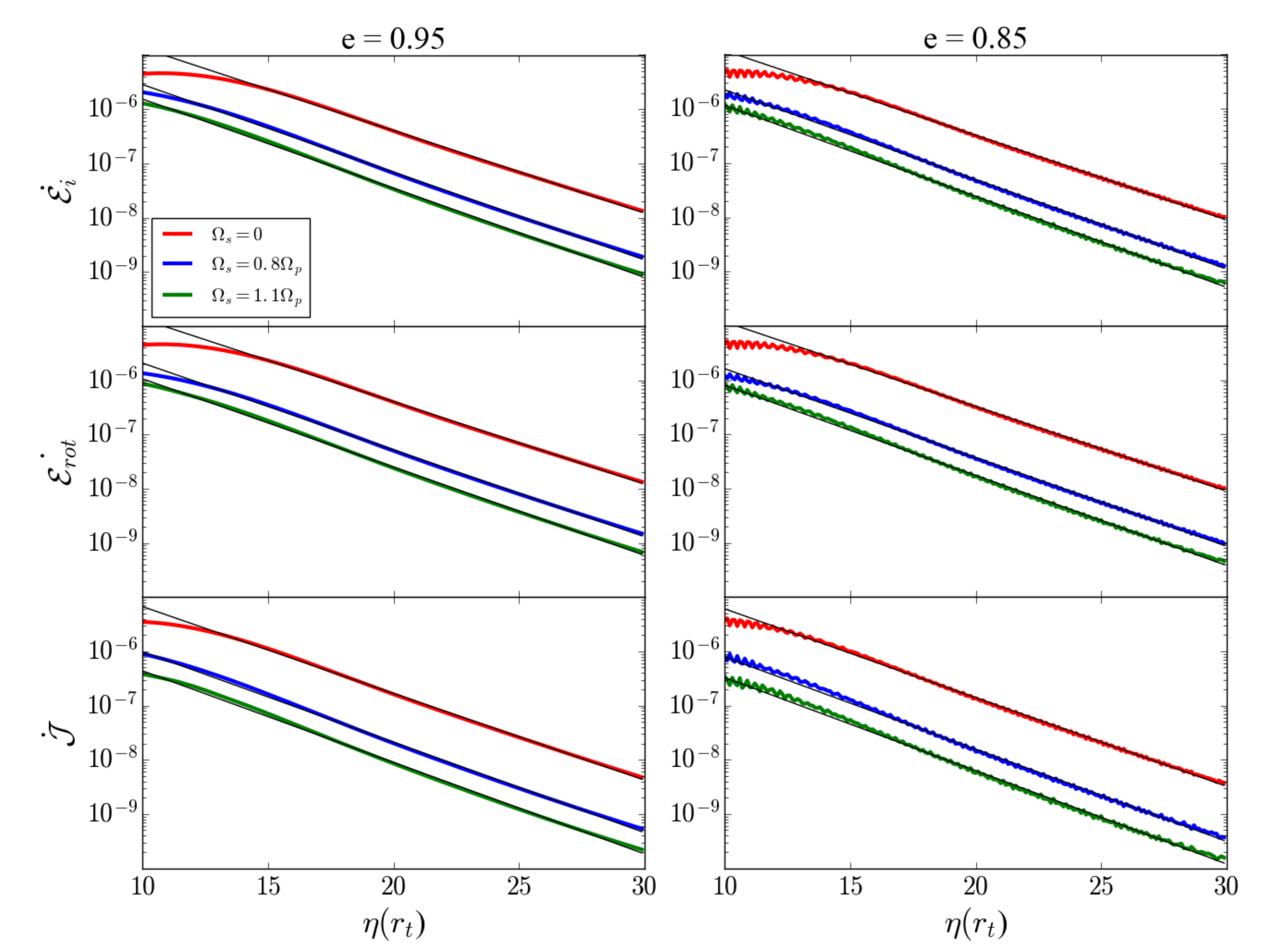}
		\caption{The dimensionless transfer rates $\dot{\mathcal{E}}_{\rm{in}}$ (on top), $\dot{\mathcal{E}}_{\text{rot}}$ (in 		the middle), and $\dot{\mathcal{J}}$ (on the bottom) as a function of $\eta$ for eccentricities of $e=0.95$ (on the left) and 		$e=0.85$ (on the right). These results are for the WD model with $T_{\rm{eff}} = 9,000$~K. Results from the older, cooler WD 			are qualitatively similar. The three colors in each panel show calculations for three values of $\Omega_{\rm{s}}$ --- 				$\Omega_{\rm{s}}=0$, and two values near $\Omega_{\rm{s}}=\Omega_{\rm{p}}$, 
	$\Omega_{\rm{s}} = 0.8\;\Omega_{\rm{p}}$ and $1.1 \;\Omega_{\rm{p}}$. Simple power law fits for $\eta \ge 15$ [ see equations (\ref{eq:Efit})-(\ref{eq:Jfit})]			are plotted in black over the numerical results. For smaller values of $\eta$, the power law fits deviate from the results. }
		\label{fig:ResultEcc95HotWDLargeEta}
	\end{figure*}

	Fig. \ref{fig:ResultEcc95HotWDLargeEta} shows some numerical results of $\dot{\mathcal{E}}_{\text{in}}$, $\dot{\mathcal{E}}_{\text{rot}}$, and $\dot{\mathcal{J}}$ in the far regime. In general, the transfer rates decrease with increasing $\eta$. This occurs because increasing $\eta$ decreases the  orbital angular velocity $\Omega$, so a term in equations (\ref{eq:dimJ})--(\ref{eq:dimErot}) with a given $N$ will correspond to a lower frequency. Because  $\hat{F}(\omega)$ is generally smaller for smaller frequencies, the overall tidal transfer rates are lower. 

	The dependence of the transfer rates on $\eta$ is stronger for lower eccentricities. This is because the Hansen coefficients $F_{Nm}$  fall off more quickly with $N$ for smaller values of $e$. As a result, fewer terms of the sums in equations (\ref{eq:dimJ})-(\ref{eq:dimErot}) contribute significantly to the transfer rates. So as the eccentricity decreases, $\dot{\mathcal{E}}_{\rm{in}}(\eta)$, $\dot{\mathcal{E}}_\text{rot}(\eta)$ , and $\dot{\mathcal{J}}(\eta)$  exhibit more of the irregular variations of the dimensionless tidal torque, $\hat{F}(\omega)$. 

	The tidal transfer rates decrease with increasing $\Omega_{\rm{s}}$. The physical reason for this is especially clear for angular momentum transfer. As the WD is spun up to a pseudo-synchronous state, the angular momentum transfer rate goes to zero. If the WD rotation rate is larger than the pseudo-synchronous rate, the transfer rate is negative (i.e., angular momentum is transferred from the WD to the orbit). The primary mathematical reason why the dimensionless transfer rates decrease with increasing $\Omega_{\rm{s}}$ is that the tidal forcing frequency is $\omega = N\Omega - 2\Omega_{\rm{s}}$. The $2\Omega_{\rm{s}}$ shift generally results in lower values of $\hat{F}(\omega)$ for a given $N$, decreasing the overall sums in equations (\ref{eq:dimJ})-(\ref{eq:dimErot}).

	 Recall that the synchronization timescale, equation (\ref{eq:tSynch}), depends inversely on $\dot{\mathcal{J}}$. Examining Fig. \ref{fig:timeScales} suggests that, for large $\eta$, the synchronization timescale will be orders of magnitude longer than the timescale for orbital evolution due to gravitational radiation. For a WD-MBH system with a large pericenter distance, it is reasonable to approximate the WD as non-rotating. In our study of the effects of tidal heating on the WD we assume a non-rotating WD. 
	
\begin{figure*}
		\includegraphics[width = 6.5in]{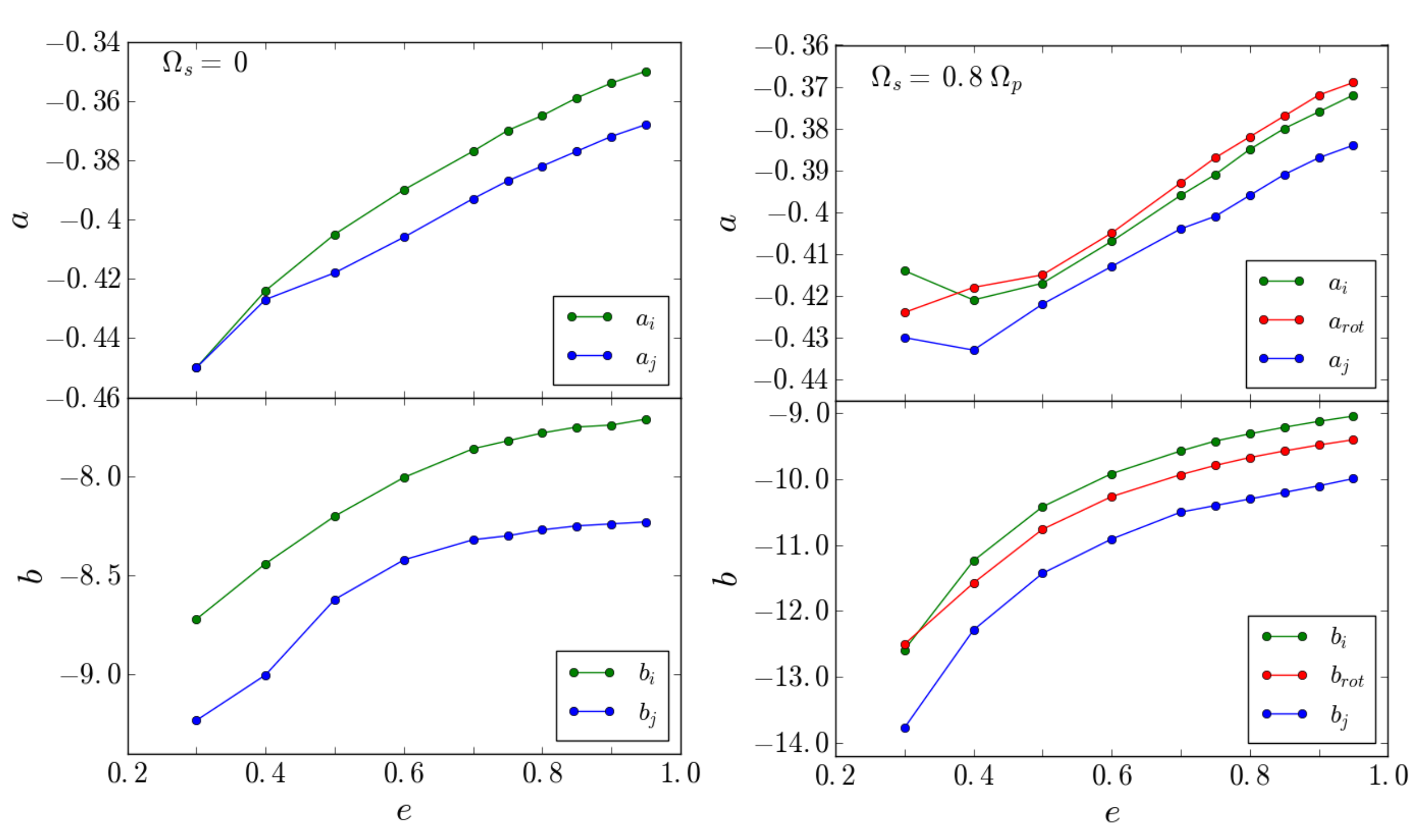}
		\caption{The parameters $a$ and $b$ found by fitting $\dot{\mathcal{E}}_{\rm{in}}(\eta)$, $\dot{\mathcal{E}}_{\rm{rot}}(\eta)$ and 
	$\dot{\mathcal{J}}(\eta)$ with equations (\ref{eq:Efit})-(\ref{eq:Jfit}) for $\Omega_{\rm{s}}=0$ (on the left) and
 	$\Omega_{\rm{s}}=0.8\;\Omega_{\rm{p}}$ (on the right) and a variety of eccentricities $e$. Both $a$ and $b$ tend to 			increase with increasing $e$. These results are for the WD model with $T_{\rm{eff}} = 9,000$~K.}
		\label{fig:abOm0}
	\end{figure*}

\begin{figure*}
		\includegraphics[width =6.5in]{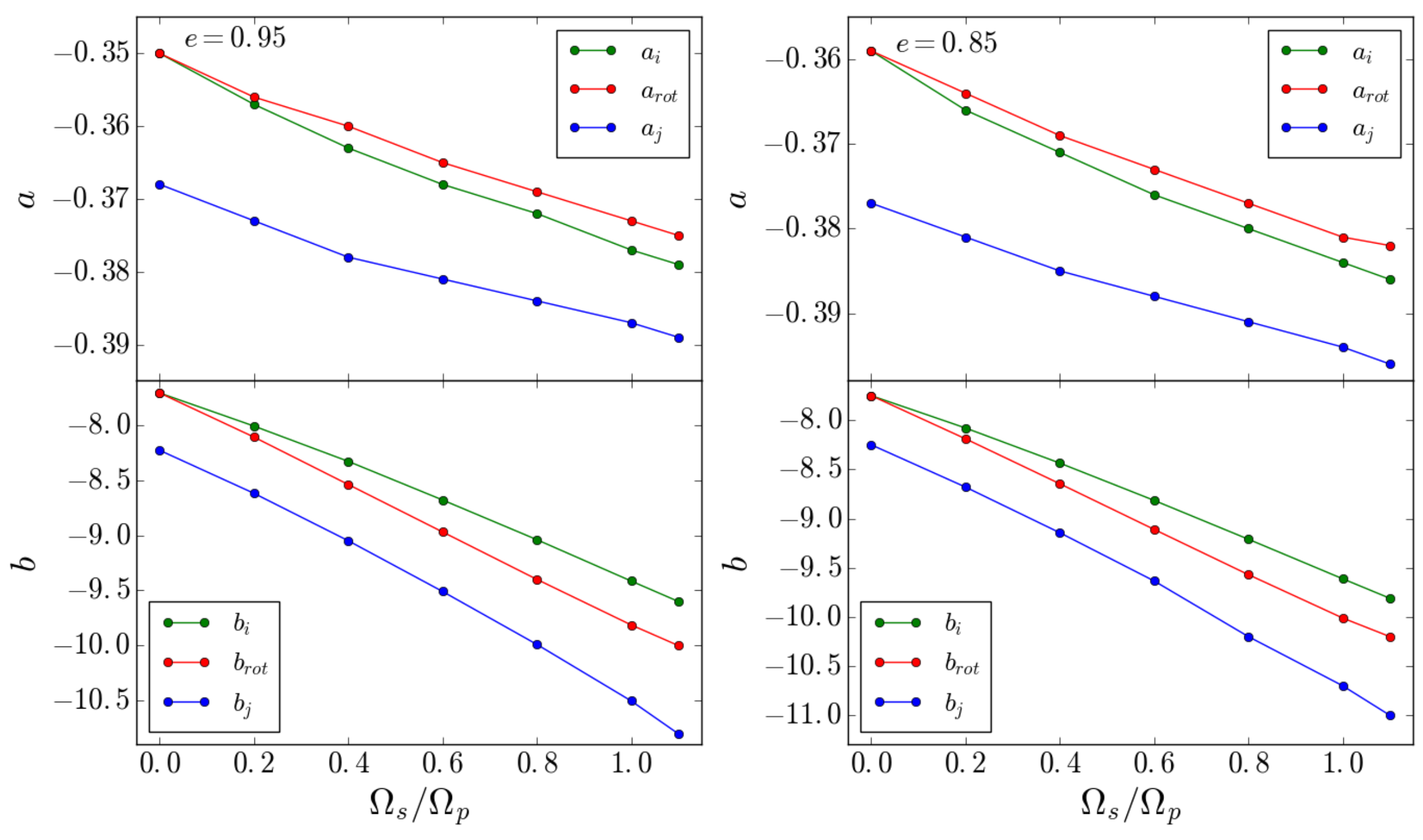}
		\caption{The parameters $a$ and $b$ found by fitting $\dot{\mathcal{E}}_{\rm{in}}(\eta)$, $\dot{\mathcal{E}}_{\rm{rot}}(\eta)$ and 
	$\dot{\mathcal{J}}(\eta)$ with equations (\ref{eq:Efit})-(\ref{eq:Jfit}) for $e=0.95$ (on the left) and $e=0.85$ (on the right) 
	and a variety of $\Omega_{\rm{s}}$. Both $a$ and $b$ tend to decrease with increasing $\Omega_{\rm{s}}$. These results 			are for the WD model with $T_{\rm{eff}} = 9,000$~K.}
		\label{fig:abEcc95}
	\end{figure*}

\begin{figure*}
		\includegraphics[width = 6.5 in]{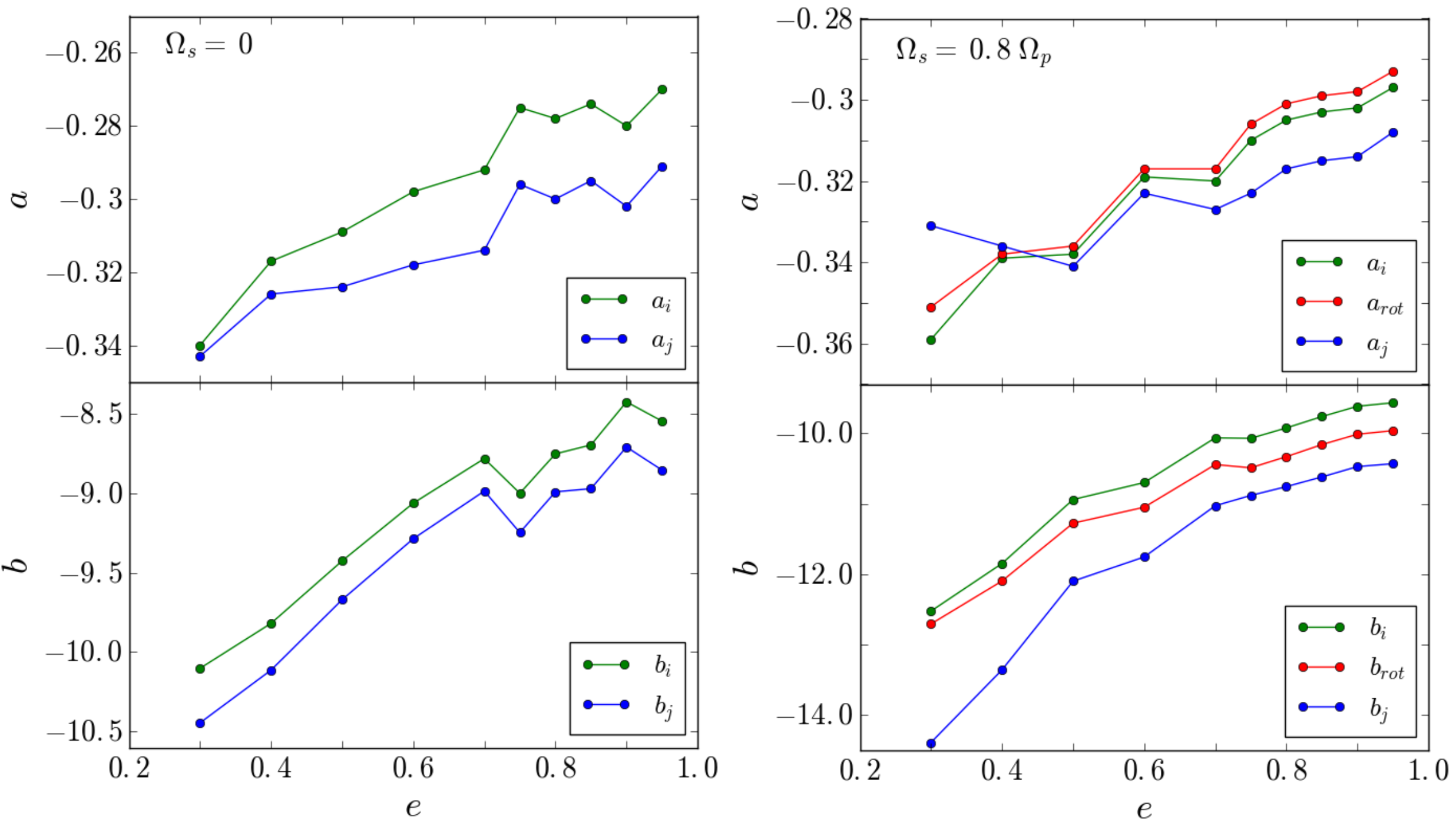}
		\caption{The parameters $a$ and $b$ found by fitting $\dot{\mathcal{E}}_{\rm{in}}(\eta)$, $\dot{\mathcal{E}}_{\rm{rot}}(\eta)$ and 
	$\dot{\mathcal{J}}(\eta)$ with equations (\ref{eq:Efit})-(\ref{eq:Jfit}) for $\Omega_{\rm{s}}=0$ (on the left) and
	$\Omega_{\rm{s}}=0.8\;\Omega_{\rm{p}}$ (on the right) and a variety of eccentricities $e$. Both $a$ and $b$ tend to 			increase with increasing $e$. These results are for the WD model with $T_{\rm{eff}} = 5,200$~K.}
		\label{fig:abOm0Cool}
	\end{figure*}

\begin{figure*}
		\includegraphics[width = 6.5in]{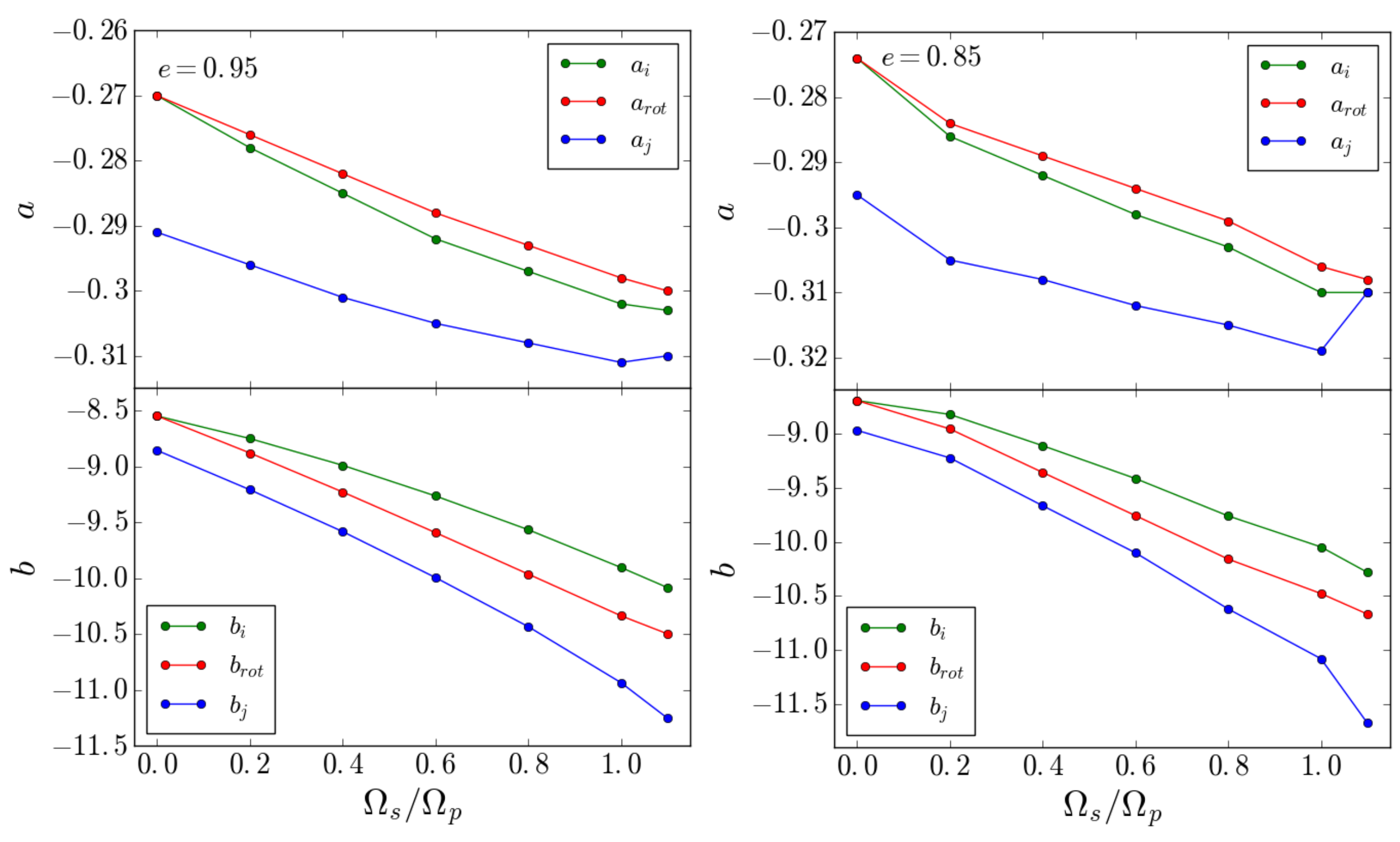}
		\caption{The parameters $a$ and $b$ found by fitting $\dot{\mathcal{E}}_{\rm{in}}(\eta)$, $\dot{\mathcal{E}}_{\rm{rot}}(\eta)$ and 
	$\dot{\mathcal{J}}(\eta)$ with equations (\ref{eq:Efit})-(\ref{eq:Jfit}) for $e=0.95$ (on the left) and $e=0.85$ (on the right) and a variety of
	$\Omega_{\rm{s}}$. Both $a$ and $b$ tend to decrease with increasing $\Omega_{\rm{s}}$. These results are for the WD model			with $T_{\rm{eff}} = 5,200$~K.}
		\label{fig:abEcc95Cool}
\end{figure*}

	Over the range $15\le \eta \le 30$, the tidal transfer rates depend on $\eta$ in a simple way, and can be fitted in the form
	\begin{align}
	\dot{\mathcal{E}}_{\text{in}}(\eta) &= \exp(a_i\eta + b_i), \label{eq:Efit}\\ 
	\dot{\mathcal{E}}_{\text{rot}}(\eta) &= \exp(a_{\rm{rot}}\eta + b_{\rm{rot}}), \label{eq:Erotfit}\\
	\dot{\mathcal{J}}(\eta) &= \exp(a_j\eta + b_j) \label{eq:Jfit}.
	\end{align}
	In the above equations, the $a$'s and $b$'s depend on $e$, $\Omega_{\rm{s}}$ and the WD models. Figs. \ref{fig:abOm0}-\ref{fig:abEcc95Cool} show the results of fittings for multiple combinations of $e$ and $\Omega_{\rm{s}}$. Both $a$'s and $b$'s tend to increase with $e$ and decrease with $\Omega_{\rm{s}}$ over some small range. The smaller the eccentricity, the more variable the transfer rates, so fittings are less certain for $e \lesssim 0.5$. In the next section, we will ignore any $\Omega_{\rm{s}}$ dependence, assuming that a WD at $\eta \gtrsim 10$ is far from synchronization and was very slowly rotating when captured.  
	%%%%%Figure%%%%%%

	%%%%%Figure%%%%%%

	%%%%%Figure%%%%%%

	%%%%%Figure%%%%%%

	%%%%%Figure%%%%%
	\begin{figure*}
		\includegraphics[width = 7in]{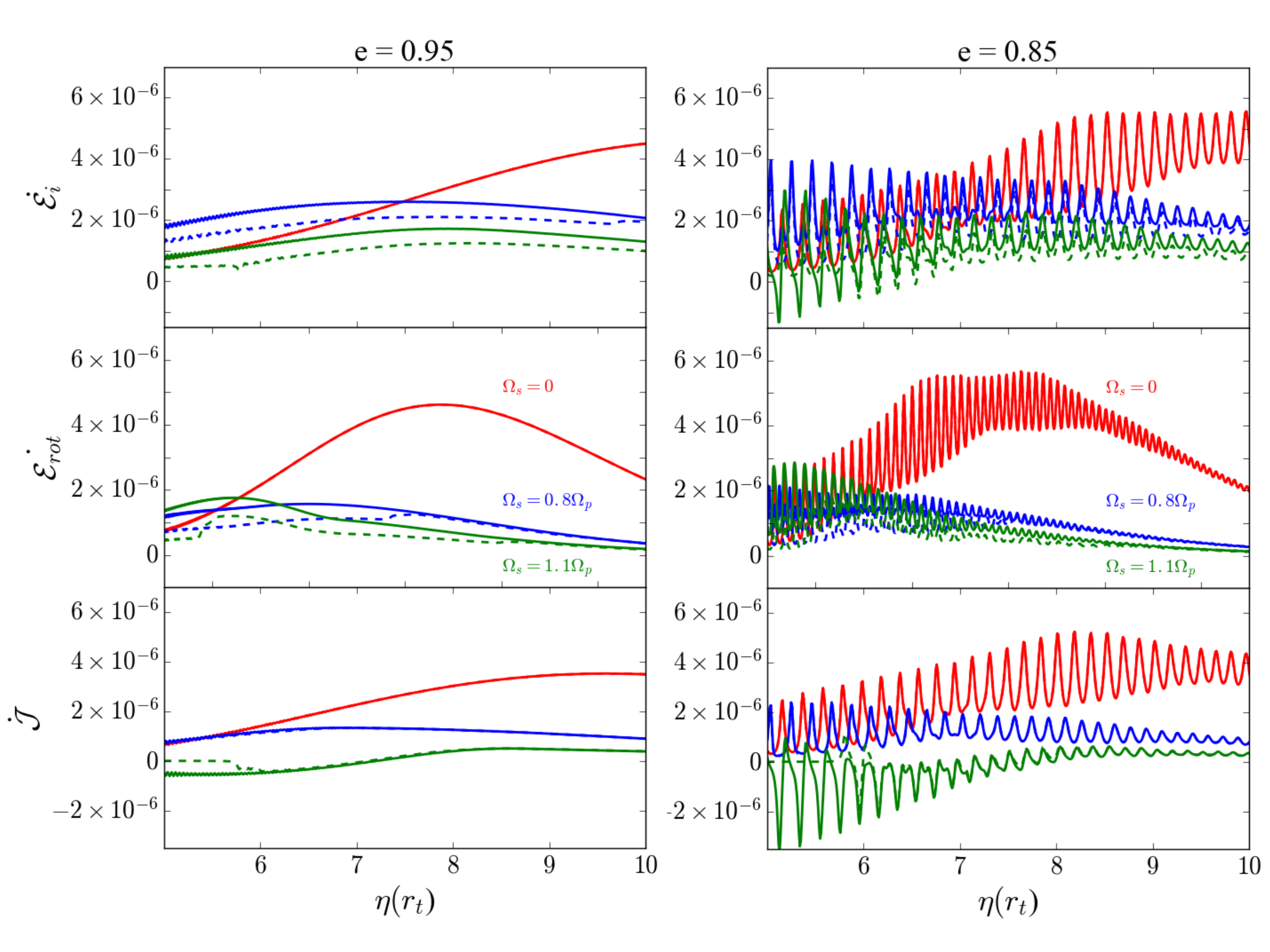}
		\caption{The dimensionless transfer rates $\dot{\mathcal{E}}_{\rm{in}}$ (on top), $\dot{\mathcal{E}}_{\text{rot}}$ (in 		the middle), and $\dot{\mathcal{J}}$ (on the bottom) as a function of $\eta$ for the WD model with 
	$T_{\rm{eff}} = 9,000$~K. 	The left and right panels show different eccentricities. The three colors in each panel show 			calculations for three values of $\Omega_{\rm{s}}$. The dotted lines show the result of excluding terms with 	
	$\omega<\Omega_{\rm{s}}$ from each sum. 	Note that the transfer rates only vary by factors of a few. In addition, 				$\dot{\mathcal{J}}$ is negative for small values of $\eta$.  }
		\label{fig:ResultEcc95HotWDSmallEta}
	\end{figure*}

	The tidal transfer rates for $\eta\le 10$ are shown in Fig. \ref{fig:ResultEcc95HotWDSmallEta}.  These results can be understood qualitatively. For the same reasons as in the far regime, the transfer rates decrease with increased $\Omega_{\rm{s}}$ and vary more for lower values of $e$. Unlike in the far regime, the transfer rates do not decrease with $\eta$. They only vary within a factor of a few. This is because, for small values of $\eta$, setting a maximum frequency of $0.1~(GM/R^3)^{1/2}$ significantly reduces the number of frequency terms that contribute to the transfer rates. For small $\eta$,  imposing the maximum frequency when $\Omega_s$ is comparable to $\Omega_{\rm{p}}$ can limit the number of positive frequency terms in equations (\ref{eq:dimJ}) and (\ref{eq:dimE}) so that $\dot{\mathcal{J}}$ and $\dot{\mathcal{E}}_{\rm{in}}$ are negative (see Fig. \ref{fig:ResultEcc95HotWDSmallEta}).  As $\eta$ increases, more terms fall within the allowed frequency range, increasing the value of the sum. This increase is counteracted by the decrease of $\hat{F}(\omega)$ with decreasing $\omega$. Calculations in this regime are less reliable than those for large $\eta$ due to a number of  reasons discussed in Section \ref{EccentricOrbit}. Additionally, calculations for low $\eta$ are more influenced by terms with $\omega < \Omega_{\rm{s}}$, where it would be necessary to consider the Coriolis force for a realistic calculation of the transfer rates. 

%%%%%%%%%%%%%%%%%%%%
\section{Tidal Heating}\label{Heating}
%%%%%%%%%%%%%%%%%%
	Though tidal dissipation has little impact on the WD's orbital evolution, tidal heating can still affect the structure of the WD. In Section \ref{Results}, we showed that for $\eta \gtrsim 15$, the tidal energy transfer rates have a simple dependence on $\eta$ [see equations (\ref{eq:Efit})-(\ref{eq:Jfit})]. In particular, $\dot{\mathcal{E}}_{\text{rot}}$ corresponds to the tidal heating rate of the WD. This allows for a first attempt at calculating the effects of tidal heating on the WD throughout orbital decay. 

\begin{figure*}
		\includegraphics[width = 4in]{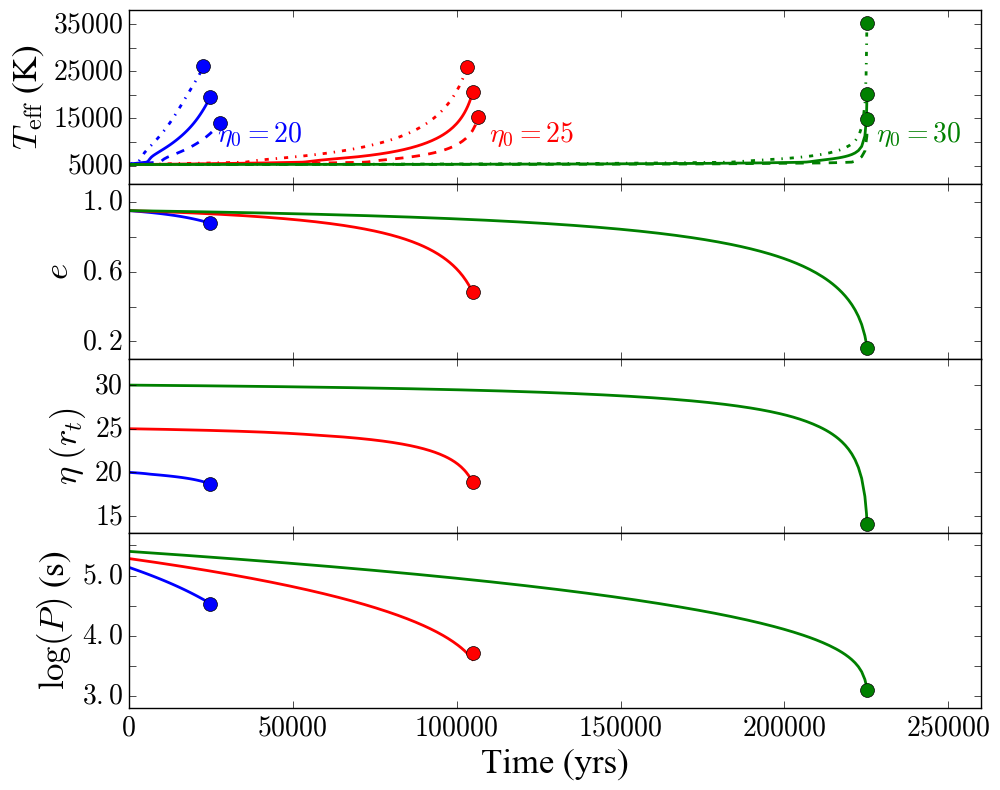}
		\caption{Evolution of the effective temperature of a MESA WD model with initial $T_{\rm{eff}}=5,200$~K orbiting a $10^5 M_\odot$ BH due to tidal heating as the orbit decays. The lower three panels show the evolution of $e$, $\eta$ and the orbital period $P$ (in seconds). The different colors represent different initial separations between the BH and WD. All systems start evolution with $e=0.95$. The solid, dashed and dotted lines show results for the envelope thicknesses $\Delta M = 10^{-4}M$, $2\times 10^{-4}M$, and $5 \times 10^{-5}M$ respectively. The circles indicate the onset of hydrogen burning in the outer envelope of the WD.}
		\label{fig:HeatingCool}
	\end{figure*}

	\begin{figure*}
		\includegraphics[width = 4in]{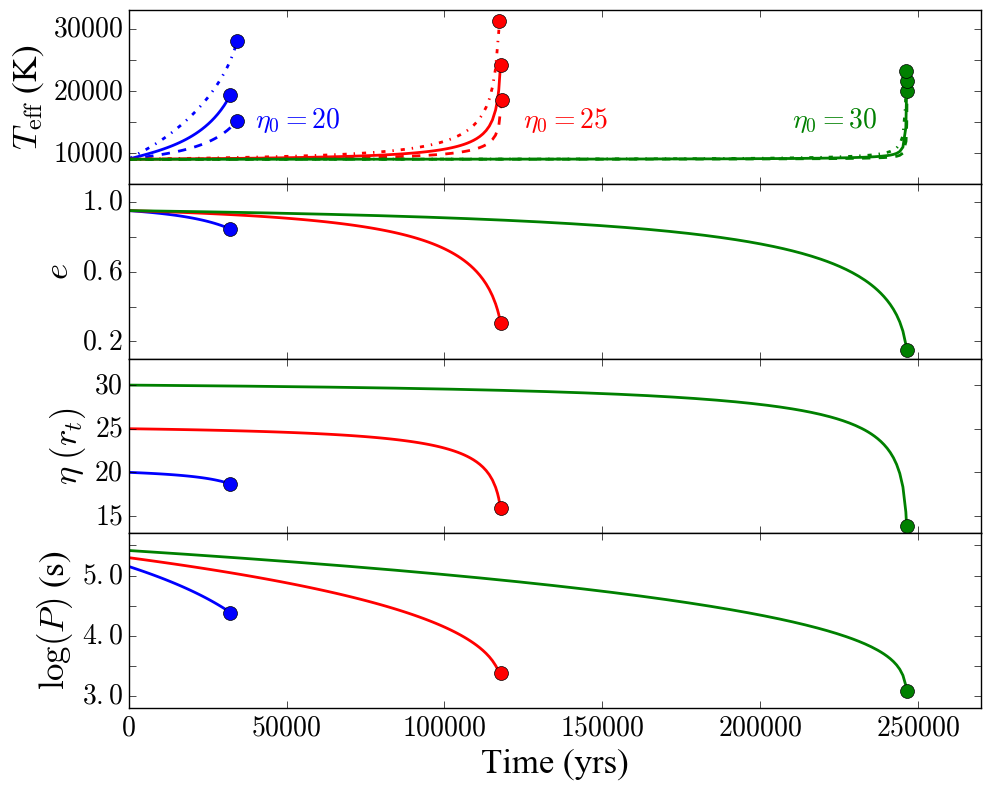}
		\caption{Same as above for a MESA WD model with initial $T_{\rm{eff}}=9,000$~K.}
		\label{fig:HeatingHot}
	\end{figure*}

	 We study the thermal evolution of the two MESA-generated WD models (described in Section \ref{DissipationCalculations}).  This calculation combines orbital evolution due to gravitational radiation and tidal heating as a function of orbital parameters $\eta$ and $e$. We approximate the WD as non-rotating, which is supported by the fact that the synchronization timescale is many times longer than the timescale for orbital evolution. When $\dot{\mathcal{E}}_{\rm{rot}}$ is known, the energy transfer rate is given by equation (\ref{eq:dimErotdef}). In terms of standard values, 
	\begin{align} \label{eq:tidalHeating}
	\dot{E}_{\rm{tide,rot}} = &1.4 \times 10^{34}~\text{erg\;s}^{-1} \left(\frac{M}{0.6 M_{\odot}}\right)^{5/2}\left(\frac{R}{0.012 R_\odot}\right)^{-5/2} \nonumber \\
	&\times \left(\frac{\eta}{10}\right)^{-15/2}  \left(\frac{\dot{\mathcal{E}}_{\rm{rot}}}{10^{-6}}\right) \left(\frac{1-e}{0.05}\right)^{3/2}.
	\end{align}
	Heat is deposited in the outer envelope of the WD as waves break via non-linear processes. For simplicity, we assume uniform heat deposition in a shell with $\Delta M \sim 10^{-4}M$. This depth corresponds to the transition from He to H in the WD, which is approximately where non-linear wave breaking occurs \citep{Fuller12c}. We inject tidal heat into the WD model and use MESA to track the profile and global properties of the WD as its orbit undergoes decay and circularization due to gravitational radiation. 

In more detail, at each timestep of orbital evolution,
	\itemize
	\item the orbital parameters $e$ and $\eta$ are integrated forward in time;
	\item new values for $a_{\rm{rot}}, a_j,b_{\rm{rot}},$ and $b_j$ are approximated by interpolating between results in Figs. \ref{fig:abOm0}-\ref{fig:abOm0Cool};
	\item $\dot{\mathcal{E}}_{\rm{rot}}$ is determined using equation (\ref{eq:Erotfit}). This value is always within a factor of a few of results from Section \ref{Results};
	\item the heating rate is calculated using equation (\ref{eq:tidalHeating}) and the assumption that heat is uniformly deposited into an envelope of depth $\Delta M$;
	\item the WD structure is evolved with the new heating rate until the next timestep using MESA.

Using this procedure, we can track the structural evolution of a WD that is captured into orbit around a BH at some initial pericenter distance $\eta_0$ and eccentricity $e_0$. The results of these calculations are shown in Figs. \ref{fig:HeatingCool} and  \ref{fig:HeatingHot}. As the WD orbit shrinks, the heating rate increases more and more rapidly. The tidal energy deposition heats the degenerate hydrogen shell and causes it to ignite (undergo thermonuclear runaway) many orbits before disruption. We used initial eccentricity $e_0=0.95$ and a variety of initial pericenter distances $\eta_0$, focusing on the evolution of systems with $\eta_0>15$ because we have the cleanest prescription for $\dot{E}_{\rm{tide,rot}}(e,\eta)$ when $\eta \gtrsim 15$.  We find that, for both WD models, the amount of heating from tides excited by a $10^5 M_\odot$ BH is enough to trigger fusion in the semi-degenerate hydrogen layer well before gravitational radiation shrinks the orbit to $\sim 3 r_{\rm{t}}$. These tidally induced novae were also suggested to occur in merging WD binaries \citep{Fuller12c} but have not been examined in the context of eccentric WD-MBH binaries.

%Changed for single spacing

\begin{figure}
		\includegraphics[width=\columnwidth]{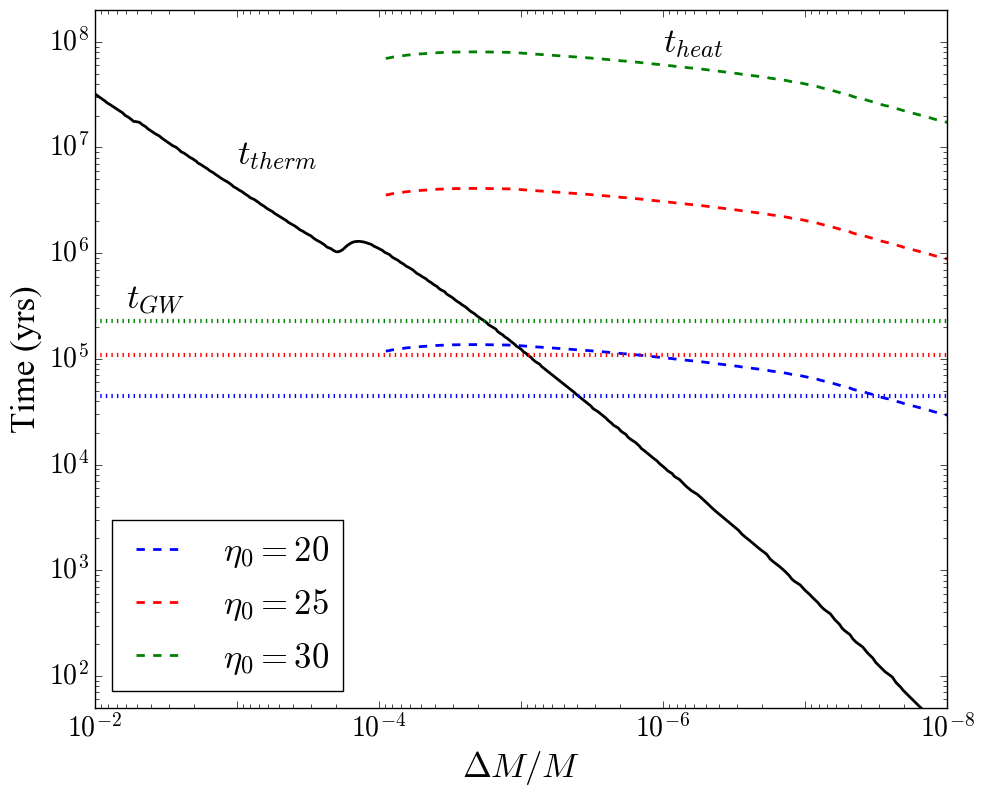}
		\caption{A comparison of timescales for the WD model with $T_{\rm{eff}}=9,000~K$ before tidal heating occurs. Envelope depth is plotted on the x-axis. The thermal timescale (solid line) characterizes the time for added heat to diffuse to the surface [see equation (\ref{eq:thermTime})]. The heating timescales (dashed lines) characterize the time for material at a given depth to heat up [see equation (\ref{eq:heatTime})]. The heating timescale is shown for heating rates at $\eta_0 = 20$, $25$ and $30$. The timescale for falling into a $10^5 M_\odot$ BH (dotted lines) is also shown for  $\eta_0 = 20$, $25$ and $30$. }
		\label{fig:HeatingTimescale}
	\end{figure}

	\begin{figure}
		\includegraphics[width =\columnwidth]{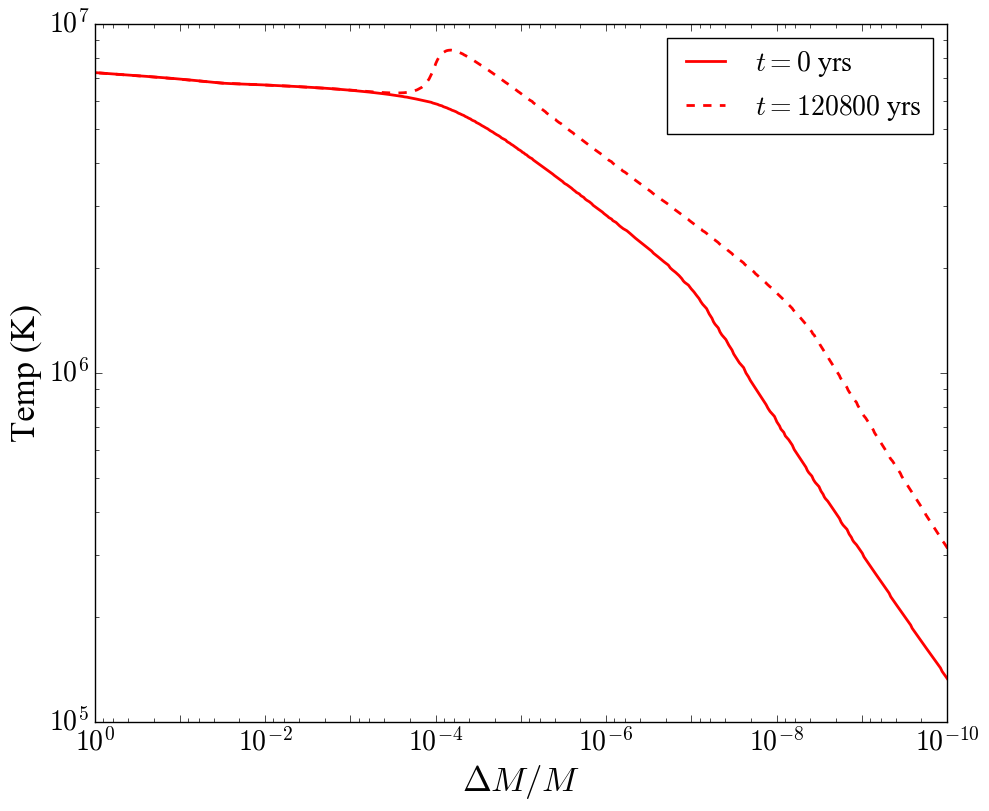}
		\caption{The temperature profile of the WD model with $T_{\rm{eff}}=9,000~K$ shown at two different times. At $t=0$ (solid line), the WD is in an orbit with $e_0=0.95$ and $\eta_0=25$. After 120,800 yrs of evolution (dashed line), the hydrogen envelope is close to ignition.}
		\label{fig:tempProfile}
	\end{figure}	

	 To understand why tidal heating induces nuclear runaway, it is useful to compare the timescale for extra heat to escape the WD with the timescale for injecting heat. The first of these is the thermal timescale $t_{\textrm{therm}}$, which characterizes the amount of time for heat to diffuse to the WD surface as a function of radius. This timescale is determined by the WD structure and given by
	\begin{equation}
	t_{\text{therm}}(r) = \frac{p c_{\rm{p}} T}{g F} \label{eq:thermTime},
	\end{equation}
	where $c_{\rm{p}}$ is the specific heat capacity at constant pressure, $p$ is pressure, $g$ is gravitational acceleration, and $F$ is energy flux. 
	Note that $t_{\text{therm}}$ varies by many orders of magnitude between the core and surface. Heat deposited in the core may take billions of years to diffuse, while heat deposited very close to the surface can exit the WD in a fraction of a year. 

	The heating timescale measures how long it takes for a given energy transfer rate to heat up of layer of material. 
	\begin{equation}
	t_{\text{heat}}(r) = \frac{c_{\rm{p}} T}{\dot{\epsilon}} \label{eq:heatTime}, 
	\end{equation}
where $\dot{\epsilon}$ is $\dot{E}_{\text{tide,rot}}/\Delta M$. Note that this timescale depends on the assumed envelope mass. A thinner envelope will both heat and cool much more quickly. The heating timescale only varies by a factor of a few across the outer envelope. When $t_{\text{heat}}\gtrsim t_{\text{therm}}$, the additional heat diffuses very quickly and in layers closest to the surface. This is the case when the WD is far away from the BH and $\dot{\mathcal{E}}_{\rm{rot}}$ is relatively small (see Fig. \ref{fig:HeatingTimescale}). However, as the WD orbit evolves, eventually $t_{\text{heat}}\lesssim t_{\text{therm}}$ at the base of the outer envelope. Then heat is trapped and the shell heats up. This is clear from examining the temperature profile of the WD just before the onset of runaway fusion (see Fig. \ref{fig:tempProfile}). Heating of the degenerate hydrogen layer can lead to runaway fusion because the pressure of this layer is independent of the temperature. For a WD captured at $\eta_0 = 20$, the base of the heating envelope already meets the criterion $t_{\text{heat}}< t_{\text{therm}}$ before gravitational radiation shrinks the orbit. In this case, the hydrogen layer quickly heats up and ignites within a few thousand years, well  before gravitational radiation drives the WD toward tidal disruption. For a WD with the same envelope captured at $\eta_0 = 30$, $t_{\text{heat}}> t_{\text{therm}}$ for a substantial fraction of the gravitational radiation timescale. In this case, heat is only trapped after the orbit has evolved long enough that $\eta$ is a about half of its original value. 

We caution that the non-linear breaking depth which determines $\Delta M$ is somewhat uncertain, as is the formation of critical layers which can absorb wave energy \citep{Fuller12c,Burkart12}. If heat deposition is limited to the outermost layers (where non-linear breaking is most assured) $t_{\rm{therm}} \ll t_{\rm{heat}}$, so the heat will diffuse out before creating a thermonuclear run away. If heat is distributed deeper in the WD, throughout the He layer, it may have little effect on the structural evolution of the WD. Assuming that non-linear breaking occurs at a critical layer of depth $\Delta M \sim 10^{-4} M$, we expect that a CO WD on the path to disruption will lose its hydrogen envelope due to a tidally induced nova before arrival.
	
\section{Summary and Discussion}
	We have studied the effects of dynamical tides on the orbital and structural evolution of a WD in a high-eccentricity orbit around a massive BH. For such WD-MBH binaries, dynamical tides involve the tidal excitation of outgoing gravity waves in the deeper envelope (around the transition region between the CO core and He layer) and their dissipation in the outer envelope of the WD.  We find that these dynamical tides have a negligible impact on the orbital evolution in comparison to gravitational radiation (GR). Additionally, the timescale for spinning up the WD via dynamical tides is generally longer than the GR-timescale. However, tidal dissipation can dramatically alter the WD structure. 

	We have calculated the rates of energy and angular momentum transfer due to tides as a function of the dimensionless pericenter distance $\eta=r_{\rm{p}}/r_{\rm{t}}$ [see equation (\ref{eq:etadef})] for various orbital eccentricities $e$ and stellar rotation rates $\Omega_{\rm{s}}$. These calculations assume that the tidally excited gravity waves are efficiently dissipated via non-linear effects or radiative damping as they propagate to the outer envelope of the WD. We have extended the method of \citet{Fuller12b}, who considered WDs in circular orbits, to eccentric orbits by decomposing the tidal potential into a sum of multiple components with different forcing frequencies and studying the WD response to each component. We have applied this method to two MESA-generated CO WD models with $T_{\rm{eff}}=9,000$ and $5,200$~K. We found that the dimensionless energy and angular momentum transfer rates, $\dot{\mathcal{E}}_{\rm{in}}(\eta)$, $\dot{\mathcal{E}}_{\rm{rot}}(\eta)$, and $\dot{\mathcal{J}}(\eta)$ [see equations (\ref{eq:dimJdef})-(\ref{eq:dimErotdef})], are relatively constant for $\eta \lesssim 10$, varying only by a factor of a few (see Fig. \ref{fig:ResultEcc95HotWDSmallEta}). The tidal transfer rates for $\eta \gtrsim 15$ decrease exponentially with increasing $\eta$ (see Fig. \ref{fig:ResultEcc95HotWDLargeEta}). For both models, increasing the stellar rotation rate $\Omega_{\rm{s}}$ decreases the transfer rates while increasing the eccentricity $e$ smooths $\dot{\mathcal{E}}_{\rm{in}}(\eta)$, $\dot{\mathcal{E}}_{\rm{rot}}(\eta)$, and $\dot{\mathcal{J}}(\eta)$. 

	We have used MESA to model the structural evolution of the WD due to tidal heating while the orbit shrinks due to gravitational radiation. We found that the hydrogen layer of the CO WD always heats up and ignites long before the orbit evolves to the point of disruption. We therefore expect that a CO WD captured into an eccentric orbit may lose its hydrogen envelope before it is torn apart by tidal forces, provided a substantial fraction of the tidal heat is deposited near the base of the hydrogen envelope.

	This paper is a first study on how dynamical tides affect a WD in an eccentric orbit around a MBH. Several caveats of our study are worth mentioning. First, the the effect of rotation is treated approximately, without including the Coriolis force. In many cases, the WD is slowly rotating and will not synchronize before disruption, so this approximation is well founded. However, a WD captured from a binary could still have a significant rotation rate. Second, we assume that the outgoing gravity waves damp efficiently near the surface. This may not be true for all tidal frequencies, in which case, our calculations overestimate tidal heating. Finally, the location of tidal heat deposition is estimated crudely and our implementation of the heating rate is approximate (especially for small $\eta$'s). 

	Our work poses the question of what happens to a WD in an eccentric orbit around a BH that has already lost its hydrogen envelope in a runaway fusion event. Future work is needed to understand how the absence of the hydrogen envelope would change both the behavior of gravity waves near the WD surface and the tidal effect at smaller
pericenter distances. 

	Our study of tidal dissipation in eccentric WD-MBH binaries can be easily adapted to other problems. For instance, WD-WD binaries with extreme eccentricities may be produced in triple systems, which could lead to direct WD-WD collisions and Type Ia supernoae (\citep{Katz12}, \citep{Kushnir13}). In such WD-WD binaries, the physics of tidal dissipation studied in this paper can be directly applied, and equations (\ref{eq:dimJdef})-(\ref{eq:dimErotdef}) remain valid. In fact, the BH mass scales out if one expresses the pericenter distance in terms of the dimensionless parameter $\eta$ [see equation (\ref{eq:tidalHeating})]. The only difference is that the orbital evolution timescale due to gravitational radiation becomes longer [see equation (\ref{eq:tGrav})] and therefore tidal dissipation may contribute to the orbital evolution.

\section*{Acknowledgments}
This work has been supported in part by NSF
grant AST-1211061, NASA grant NNX14AP31G
and a Simons fellowship from the Simons Foundation. 
MV is supported by a NASA Earth and Space Sciences Fellowship in Astrophysics.

\bibliographystyle{mnras}	
\bibliography{References}

\begin{thebibliography}{}
\makeatletter
\relax
\def\mn@urlcharsother{\let\do\@makeother \do\$\do\&\do\#\do\^\do\_\do\%\do\~}
\def\mn@doi{\begingroup\mn@urlcharsother \@ifnextchar [ {\mn@doi@}
  {\mn@doi@[]}}
\def\mn@doi@[#1]#2{\def\@tempa{#1}\ifx\@tempa\@empty \href
  {http://dx.doi.org/#2} {doi:#2}\else \href {http://dx.doi.org/#2} {#1}\fi
  \endgroup}
\def\mn@eprint#1#2{\mn@eprint@#1:#2::\@nil}
\def\mn@eprint@arXiv#1{\href {http://arxiv.org/abs/#1} {{\tt arXiv:#1}}}
\def\mn@eprint@dblp#1{\href {http://dblp.uni-trier.de/rec/bibtex/#1.xml}
  {dblp:#1}}
\def\mn@eprint@#1:#2:#3:#4\@nil{\def\@tempa {#1}\def\@tempb {#2}\def\@tempc
  {#3}\ifx \@tempc \@empty \let \@tempc \@tempb \let \@tempb \@tempa \fi \ifx
  \@tempb \@empty \def\@tempb {arXiv}\fi \@ifundefined
  {mn@eprint@\@tempb}{\@tempb:\@tempc}{\expandafter \expandafter \csname
  mn@eprint@\@tempb\endcsname \expandafter{\@tempc}}}

\bibitem[\protect\citeauthoryear{{Barker} \& {Ogilvie}}{{Barker} \&
  {Ogilvie}}{2010}]{Barker10}
{Barker} A.~J.,  {Ogilvie} G.~I.,  2010, \mn@doi [\mnras]
  {10.1111/j.1365-2966.2010.16400.x}, \href
  {http://adsabs.harvard.edu/abs/2010MNRAS.404.1849B} {404, 1849}

\bibitem[\protect\citeauthoryear{{Barker} \& {Ogilvie}}{{Barker} \&
  {Ogilvie}}{2011}]{Barker11}
{Barker} A.~J.,  {Ogilvie} G.~I.,  2011, \mn@doi [\mnras]
  {10.1111/j.1365-2966.2011.19322.x}, \href
  {http://adsabs.harvard.edu/abs/2011MNRAS.417..745B} {417, 745}

\bibitem[\protect\citeauthoryear{{Bloom} et~al.,}{{Bloom}
  et~al.}{2011}]{Bloom11}
{Bloom} J.~S.,  et~al., 2011, \mn@doi [Science] {10.1126/science.1207150},
  \href {http://adsabs.harvard.edu/abs/2011Sci...333..203B} {333, 203}

\bibitem[\protect\citeauthoryear{{Burkart}, {Quataert}, {Arras}  \&
  {Weinberg}}{{Burkart} et~al.}{2012}]{Burkart12}
{Burkart} J.,  {Quataert} E.,  {Arras} P.,   {Weinberg} N.~N.,  2012, \mn@doi
  [\mnras] {10.1111/j.1365-2966.2011.20344.x}, \href
  {http://adsabs.harvard.edu/abs/2012MNRAS.421..983B} {421, 983}

\bibitem[\protect\citeauthoryear{{Burrows} et~al.,}{{Burrows}
  et~al.}{2011}]{Burrows11}
{Burrows} D.~N.,  et~al., 2011, \mn@doi [\nat] {10.1038/nature10374}, \href
  {http://adsabs.harvard.edu/abs/2011Natur.476..421B} {476, 421}

\bibitem[\protect\citeauthoryear{{Cheng} \& {Bogdanovi{\'c}}}{{Cheng} \&
  {Bogdanovi{\'c}}}{2014}]{Cheng14}
{Cheng} R.~M.,  {Bogdanovi{\'c}} T.,  2014, \mn@doi [\prd]
  {10.1103/PhysRevD.90.064020}, \href
  {http://adsabs.harvard.edu/abs/2014PhRvD..90f4020C} {90, 064020}

\bibitem[\protect\citeauthoryear{{Cheng} \& {Evans}}{{Cheng} \&
  {Evans}}{2013}]{Cheng13}
{Cheng} R.~M.,  {Evans} C.~R.,  2013, \mn@doi [\prd]
  {10.1103/PhysRevD.87.104010}, \href
  {http://adsabs.harvard.edu/abs/2013PhRvD..87j4010C} {87, 104010}

\bibitem[\protect\citeauthoryear{{De Colle}, {Guillochon}, {Naiman}  \&
  {Ramirez-Ruiz}}{{De Colle} et~al.}{2012}]{Decolle12}
{De Colle} F.,  {Guillochon} J.,  {Naiman} J.,   {Ramirez-Ruiz} E.,  2012,
  \mn@doi [\apj] {10.1088/0004-637X/760/2/103}, \href
  {http://adsabs.harvard.edu/abs/2012ApJ...760..103D} {760, 103}

\bibitem[\protect\citeauthoryear{{East}}{{East}}{2014}]{East14}
{East} W.~E.,  2014, \mn@doi [\apj] {10.1088/0004-637X/795/2/135}, \href
  {http://adsabs.harvard.edu/abs/2014ApJ...795..135E} {795, 135}

\bibitem[\protect\citeauthoryear{{Fuller} \& {Lai}}{{Fuller} \&
  {Lai}}{2012a}]{Fuller12b}
{Fuller} J.,  {Lai} D.,  2012a, \mn@doi [\mnras]
  {10.1111/j.1365-2966.2011.20320.x}, \href
  {http://adsabs.harvard.edu/abs/2012MNRAS.421..426F} {421, 426}

\bibitem[\protect\citeauthoryear{{Fuller} \& {Lai}}{{Fuller} \&
  {Lai}}{2012b}]{Fuller12c}
{Fuller} J.,  {Lai} D.,  2012b, \mn@doi [\apjl] {10.1088/2041-8205/756/1/L17},
  \href {http://adsabs.harvard.edu/abs/2012ApJ...756L..17F} {756, L17}

\bibitem[\protect\citeauthoryear{{Fuller} \& {Lai}}{{Fuller} \&
  {Lai}}{2013}]{Fuller13}
{Fuller} J.,  {Lai} D.,  2013, \mn@doi [\mnras] {10.1093/mnras/sts606}, \href
  {http://adsabs.harvard.edu/abs/2013MNRAS.430..274F} {430, 274}

\bibitem[\protect\citeauthoryear{{Fuller} \& {Lai}}{{Fuller} \&
  {Lai}}{2014}]{Fuller14}
{Fuller} J.,  {Lai} D.,  2014, \mn@doi [\mnras] {10.1093/mnras/stu1698}, \href
  {http://adsabs.harvard.edu/abs/2014MNRAS.444.3488F} {444, 3488}

\bibitem[\protect\citeauthoryear{{Giannios} \& {Metzger}}{{Giannios} \&
  {Metzger}}{2011}]{Giannios11}
{Giannios} D.,  {Metzger} B.~D.,  2011, \mn@doi [\mnras]
  {10.1111/j.1365-2966.2011.19188.x}, \href
  {http://adsabs.harvard.edu/abs/2011MNRAS.416.2102G} {416, 2102}

\bibitem[\protect\citeauthoryear{{Goldreich} \& {Nicholson}}{{Goldreich} \&
  {Nicholson}}{1989}]{Goldreich89}
{Goldreich} P.,  {Nicholson} P.~D.,  1989, \mn@doi [\apj] {10.1086/167665},
  \href {http://adsabs.harvard.edu/abs/1989ApJ...342.1079G} {342, 1079}

\bibitem[\protect\citeauthoryear{{Goodman} \& {Dickson}}{{Goodman} \&
  {Dickson}}{1998}]{Goodman98}
{Goodman} J.,  {Dickson} E.~S.,  1998, \mn@doi [\apj] {10.1086/306348}, \href
  {http://adsabs.harvard.edu/abs/1998ApJ...507..938G} {507, 938}

\bibitem[\protect\citeauthoryear{{Hills}}{{Hills}}{1975}]{Hills75}
{Hills} J.~G.,  1975, \mn@doi [\nat] {10.1038/254295a0}, \href
  {http://adsabs.harvard.edu/abs/1975Natur.254..295H} {254, 295}

\bibitem[\protect\citeauthoryear{{Ioka}, {Hotokezaka}  \& {Piran}}{{Ioka}
  et~al.}{2016}]{Ioka16}
{Ioka} K.,  {Hotokezaka} K.,   {Piran} T.,  2016, preprint, \href
  {http://adsabs.harvard.edu/abs/2016arXiv160802938I} {} (\mn@eprint {arXiv}
  {1608.02938})

\bibitem[\protect\citeauthoryear{{Katz} \& {Dong}}{{Katz} \&
  {Dong}}{2012}]{Katz12}
{Katz} B.,  {Dong} S.,  2012, preprint, \href
  {http://adsabs.harvard.edu/abs/2012arXiv1211.4584K} {} (\mn@eprint {arXiv}
  {1211.4584})

\bibitem[\protect\citeauthoryear{{Krolik} \& {Piran}}{{Krolik} \&
  {Piran}}{2012}]{Krolik12}
{Krolik} J.~H.,  {Piran} T.,  2012, \mn@doi [\apj]
  {10.1088/0004-637X/749/1/92}, \href
  {http://adsabs.harvard.edu/abs/2012ApJ...749...92K} {749, 92}

\bibitem[\protect\citeauthoryear{{Kushnir}, {Katz}, {Dong}, {Livne}  \&
  {Fern{\'a}ndez}}{{Kushnir} et~al.}{2013}]{Kushnir13}
{Kushnir} D.,  {Katz} B.,  {Dong} S.,  {Livne} E.,   {Fern{\'a}ndez} R.,  2013,
  \mn@doi [\apjl] {10.1088/2041-8205/778/2/L37}, \href
  {http://adsabs.harvard.edu/abs/2013ApJ...778L..37K} {778, L37}

\bibitem[\protect\citeauthoryear{{Lai}}{{Lai}}{1997}]{Lai97}
{Lai} D.,  1997, \apj, \href
  {http://adsabs.harvard.edu/abs/1997ApJ...490..847L} {490, 847}

\bibitem[\protect\citeauthoryear{{Levan} et~al.,}{{Levan}
  et~al.}{2014}]{Levan14}
{Levan} A.~J.,  et~al., 2014, \mn@doi [\apj] {10.1088/0004-637X/781/1/13},
  \href {http://adsabs.harvard.edu/abs/2014ApJ...781...13L} {781, 13}

\bibitem[\protect\citeauthoryear{{Luminet} \& {Pichon}}{{Luminet} \&
  {Pichon}}{1989}]{Luminet89}
{Luminet} J.-P.,  {Pichon} B.,  1989, \aap, \href
  {http://adsabs.harvard.edu/abs/1989A%26A...209..103L} {209, 103}

\bibitem[\protect\citeauthoryear{{MacLeod}, {Goldstein}, {Ramirez-Ruiz},
  {Guillochon}  \& {Samsing}}{{MacLeod} et~al.}{2014}]{MacLeod14}
{MacLeod} M.,  {Goldstein} J.,  {Ramirez-Ruiz} E.,  {Guillochon} J.,
  {Samsing} J.,  2014, \mn@doi [\apj] {10.1088/0004-637X/794/1/9}, \href
  {http://adsabs.harvard.edu/abs/2014ApJ...794....9M} {794, 9}

\bibitem[\protect\citeauthoryear{{MacLeod}, {Guillochon}, {Ramirez-Ruiz},
  {Kasen}  \& {Rosswog}}{{MacLeod} et~al.}{2016}]{MacLeod16}
{MacLeod} M.,  {Guillochon} J.,  {Ramirez-Ruiz} E.,  {Kasen} D.,   {Rosswog}
  S.,  2016, \mn@doi [\apj] {10.3847/0004-637X/819/1/3}, \href
  {http://adsabs.harvard.edu/abs/2016ApJ...819....3M} {819, 3}

\bibitem[\protect\citeauthoryear{Murray \& Dermott}{Murray \&
  Dermott}{2000}]{Murray00}
Murray C.~D.,  Dermott S.,  2000, Solar System Dynamics.
Cambridge Univ. Press, Cambridge

\bibitem[\protect\citeauthoryear{{Ogilvie} \& {Lin}}{{Ogilvie} \&
  {Lin}}{2007}]{Ogilvie07}
{Ogilvie} G.~I.,  {Lin} D.~N.~C.,  2007, \mn@doi [\apj] {10.1086/515435}, \href
  {http://adsabs.harvard.edu/abs/2007ApJ...661.1180O} {661, 1180}

\bibitem[\protect\citeauthoryear{{Paxton}, {Bildsten}, {Dotter}, {Herwig},
  {Lesaffre}  \& {Timmes}}{{Paxton} et~al.}{2011}]{Paxton11}
{Paxton} B.,  {Bildsten} L.,  {Dotter} A.,  {Herwig} F.,  {Lesaffre} P.,
  {Timmes} F.,  2011, \mn@doi [\apjs] {10.1088/0067-0049/192/1/3}, \href
  {http://adsabs.harvard.edu/abs/2011ApJS..192....3P} {192, 3}

\bibitem[\protect\citeauthoryear{Peters}{Peters}{1964}]{Peters64}
Peters P.~C.,  1964, Phys. Rev., 136

\bibitem[\protect\citeauthoryear{{Press} \& {Teukolsky}}{{Press} \&
  {Teukolsky}}{1977}]{Press77}
{Press} W.~H.,  {Teukolsky} S.~A.,  1977, \mn@doi [\apj] {10.1086/155143},
  \href {http://adsabs.harvard.edu/abs/1977ApJ...213..183P} {213, 183}

\bibitem[\protect\citeauthoryear{{Rees}}{{Rees}}{1988}]{Rees88}
{Rees} M.~J.,  1988, \mn@doi [\nat] {10.1038/333523a0}, \href
  {http://adsabs.harvard.edu/abs/1988Natur.333..523R} {333, 523}

\bibitem[\protect\citeauthoryear{{Rosswog}, {Ramirez-Ruiz}, {Hix}  \&
  {Dan}}{{Rosswog} et~al.}{2008a}]{Rosswog08b}
{Rosswog} S.,  {Ramirez-Ruiz} E.,  {Hix} W.~R.,   {Dan} M.,  2008a, \mn@doi
  [Computer Physics Communications] {10.1016/j.cpc.2008.01.031}, \href
  {http://adsabs.harvard.edu/abs/2008CoPhC.179..184R} {179, 184}

\bibitem[\protect\citeauthoryear{{Rosswog}, {Ramirez-Ruiz}  \& {Hix}}{{Rosswog}
  et~al.}{2008b}]{Rosswog08a}
{Rosswog} S.,  {Ramirez-Ruiz} E.,   {Hix} W.~R.,  2008b, \mn@doi [\apj]
  {10.1086/528738}, \href {http://adsabs.harvard.edu/abs/2008ApJ...679.1385R}
  {679, 1385}

\bibitem[\protect\citeauthoryear{{Rosswog}, {Ramirez-Ruiz}  \& {Hix}}{{Rosswog}
  et~al.}{2009}]{Rosswog09}
{Rosswog} S.,  {Ramirez-Ruiz} E.,   {Hix} W.~R.,  2009, \mn@doi [\apj]
  {10.1088/0004-637X/695/1/404}, \href
  {http://adsabs.harvard.edu/abs/2009ApJ...695..404R} {695, 404}

\bibitem[\protect\citeauthoryear{{Sesana}, {Vecchio}, {Eracleous}  \&
  {Sigurdsson}}{{Sesana} et~al.}{2008}]{Sesana08}
{Sesana} A.,  {Vecchio} A.,  {Eracleous} M.,   {Sigurdsson} S.,  2008, \mn@doi
  [\mnras] {10.1111/j.1365-2966.2008.13904.x}, \href
  {http://adsabs.harvard.edu/abs/2008MNRAS.391..718S} {391, 718}

\bibitem[\protect\citeauthoryear{{Shcherbakov}, {Pe'er}, {Reynolds}, {Haas},
  {Bode}  \& {Laguna}}{{Shcherbakov} et~al.}{2013}]{Shcherbakov13}
{Shcherbakov} R.~V.,  {Pe'er} A.,  {Reynolds} C.~S.,  {Haas} R.,  {Bode} T.,
  {Laguna} P.,  2013, \mn@doi [\apj] {10.1088/0004-637X/769/2/85}, \href
  {http://adsabs.harvard.edu/abs/2013ApJ...769...85S} {769, 85}

\bibitem[\protect\citeauthoryear{{Zahn}}{{Zahn}}{1975}]{Zahn75}
{Zahn} J.-P.,  1975, \aap, \href
  {http://adsabs.harvard.edu/abs/1975A%26A....41..329Z} {41, 329}

\bibitem[\protect\citeauthoryear{{Zalamea}, {Menou}  \&
  {Beloborodov}}{{Zalamea} et~al.}{2010}]{Zalamea10}
{Zalamea} I.,  {Menou} K.,   {Beloborodov} A.~M.,  2010, \mn@doi [\mnras]
  {10.1111/j.1745-3933.2010.00930.x}, \href
  {http://adsabs.harvard.edu/abs/2010MNRAS.409L..25Z} {409, L25}

\bibitem[\protect\citeauthoryear{{van Velzen} et~al.,}{{van Velzen}
  et~al.}{2016}]{vanVelzen16}
{van Velzen} S.,  et~al., 2016, \mn@doi [Science] {10.1126/science.aad1182},
  \href {http://adsabs.harvard.edu/abs/2016Sci...351...62V} {351, 62}

\makeatother
\end{thebibliography}

\appendix
\section{Hansen Coefficients}\label{sec:ApA}
 	For large $\eta$ and $e$ it is often necessary to calculate the Hansen coefficient $F_{Nm}$, for $N\gg1$ to accurately determine the tidal energy and angular momentum transfer rates. For illustration, consider a system with $\eta=20$, $e=0.95$ and $\Omega_{\rm{s}}=0$.  The largest $\omega$ included in our calculations is $\omega \sim 0.1~(GM/R^3)^{1/2}$ because gravity waves with larger $\omega$ will not propagate in the WD (see Section \ref{EccentricOrbit}). Therefore, the largest $N$ that appears in equations (\ref{eq:Jrate})-(\ref{eq:Erotrate}) is given by
\begin{equation}
	N_{\text{max}} = \Bigl\lfloor0.1 \left(\frac{1-e}{\eta}\right)^{3/2}\Bigr\rfloor.
\end{equation}
For the system described above, $N_{\text{max}} = 800$. Note that although $F_{Nm}$ decreases with increasing $N$, the dimensionless tidal torque $\hat{F}(\omega)$ increases very steeply with $\omega$. These two effects can balance each other so that terms with large $N$ contribute significantly to the tidal transfer rates, equations (\ref{eq:Jrate})-(\ref{eq:Erotrate}).

	In order to calculate $F_{Nm}$ for large $N$ and $e$, it is useful to find an approximation. These Hansen coefficients are calculated by integrating over an oscillatory term [see equation (\ref{eq:HansenCoeff})]. When $N$ is large, there are many oscillations over the interval of integration, so numerical integration is inefficient and, unless handled carefully, inaccurate. We therefore treat the $F_{Nm}$ as a continuous function, $F_m(N)$ and fit this function with the form:
\begin{equation}
	F_m(N) = \alpha N^{\beta} \rm{exp}(-\gamma N).
	\label{eq:HansenApprox}
\end{equation}
This form is motivated by approximations for the parabolic case (e=1) in \citet{Press77} and \citet{Lai97}. 

\begin{figure*}
	\includegraphics[width=5 in]{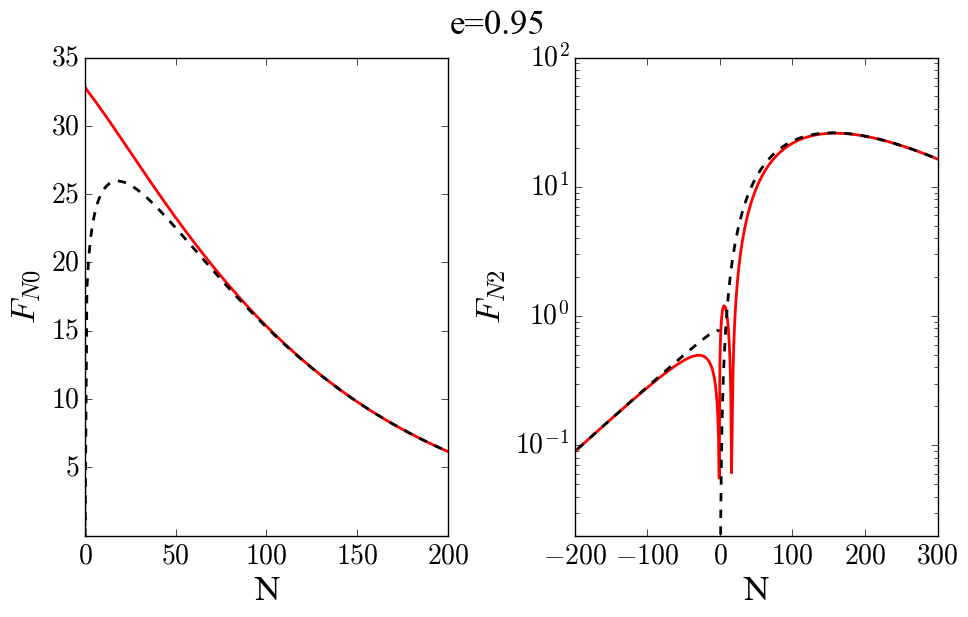}
	\caption{The Hansen coefficients $F_{Nm}$ in solid red and an approximation of the form equation (\ref{eq:HansenApprox}) in dashed black. Both panels use $e=0.95$. The left panel shows results for $m=0$ while the right panel shows results for $m=2$. From the symmetry of the Hansen coefficients, $F_{N2} = F_{-N-2}$. For large $N$, the approximation is accurate to $0.01\%$. }
	\label{fig:Hansen}
\end{figure*}

Fig. \ref{fig:Hansen} shows a comparison between $F_{Nm}$ and the approximation $F_m(N)$ for $e=0.95$. With that eccentricity, 
	\begin{align}
	F_0(N) &\approx 18.3 \: N^{0.186} \rm{exp}(-0.0103\;N), \nonumber\\
	F_2(N) &\approx 0.024 \:N^{1.73}\rm{exp}(-0.0111\;N), \nonumber \\
	F_{-2}(N)&\approx 0.766 \:N^{0.040}\rm{exp}(-0.0118\;N).
	\end{align}
For large $e$ the approximation is accurate to within $0.01\%$.

\bsp	% typesetting comment
\label{lastpage}

\end{document}